\documentclass[sigconf]{acmart}
\usepackage{booktabs} % For formal tables
\usepackage{multirow}
\usepackage{subfig}
\usepackage{listings}
\usepackage{xspace}
\usepackage{tabularx}
\usepackage{url}
\usepackage{amssymb}
\usepackage{wasysym}
\usepackage{xcolor,color, colortbl}%\usepackage{proof}
\usepackage{pifont}
\usepackage{wrapfig}
\usepackage{adjustbox}
\usepackage[normalem]{ulem}
\usepackage{algorithmic}
\usepackage[linesnumbered,ruled,vlined]{algorithm2e}
\usepackage{stmaryrd}

\usepackage{caption}

\newcommand{\distance}{3pt}
\begin{document}
%\title{Simplistic Program Repair}
\title{Practical Program Repair via Bytecode Mutation}

\author{Ali Ghanbari and Lingming Zhang}
\affiliation{
  \institution{The University of Texas at Dallas}
  %\streetaddress{800 W Campbell Rd}
  %\city{Richardson}
  %\state{Texas}
  %\postcode{75080-1558}
}
\email{{ali.ghanbari,  lingming.zhang}@utdallas.edu}

%util
\newcommand{\Comment}[1]{}
\newcommand{\todo}[1]{\textcolor[rgb]{1.0,0.0,0.0}{#1}}
\newcommand{\lingming}[1]{\textcolor[rgb]{1.0,0.0,0.0}{#1}}
\newcommand{\ali}[1]{\textcolor[rgb]{0.0,0.0,1.0}{#1}}
\definecolor{dkgreen}{rgb}{0,0.6,0}
\definecolor{GrayOne}{gray}{0.9}
\definecolor{GrayTwo}{gray}{0.8}
\definecolor{GrayThree}{gray}{0.7}
\definecolor{GrayFour}{gray}{0.6}
\definecolor{GrayFive}{gray}{0.5}

\lstset{escapeinside={*@}{@*},}
\lstset{language=Java,
        columns=fullflexible,
        basicstyle={\small\ttfamily},
        keywordstyle=\bfseries,
        %stringstyle=\color{Green},
        commentstyle=\color{gray},
        morecomment=[s][\color{javadocblue}]{/**}{*/},
}
\newcommand{\codeIn}[1]{\lstinline!#1!}
\newcommand{\parabf}[1]{\noindent \textbf{#1}}
%techs
\newcommand{\simpr}{PraPR\xspace}
\newcommand{\pit}{PIT\xspace}
\newcommand{\apr}{APR\xspace}
\newcommand{\defectsj}{Defects4J\xspace}

\newcommand{\capGen}{CapGen\xspace}
\newcommand{\jaid}{JAID\xspace}
\newcommand{\elixir}{ELIXIR\xspace}
\newcommand{\acs}{ACS\xspace}
\newcommand{\hdRepair}{HD-Repair\xspace}
\newcommand{\xpar}{xPAR\xspace}
\newcommand{\nopol}{NOPOL\xspace}
\newcommand{\jGenProg}{jGenProg\xspace}
\newcommand{\jMut}{jMutRepair\xspace}
\newcommand{\jKali}{jKali\xspace}

%numbers
\newcommand{\djbug}{395\xspace}
\newcommand{\ppatch}{113\xspace} %148 total
\newcommand{\gpatch}{18\xspace} %43 total
\newcommand{\allgenuines}{43\xspace}

%symbols

\newcommand{\defn}{de\!f\!n}
\newcommand{\field}{f\!ield}
\newcommand{\meth}{meth}
\newcommand{\body}{body}
\newcommand{\argdec}{arg}
\newcommand{\var}{var}
\newcommand{\md}{md}
\newcommand{\fd}{f\!d}
\newcommand{\are}{ae}
\newcommand{\be}{be}
\newcommand{\cjmath}[1]{\mathsf{\mathbf{#1}}}
\newcommand{\cst}{ct}
\newcommand{\defval}{\mathrm{defVal}}
\newcommand{\vlentry}{V\!L_{\mathrm{entry}}}
\newcommand{\vlexit}{V\!L_{\mathrm{exit}}}
\newcommand{\gen}{\mathrm{gen}}
\newcommand{\killl}{\mathrm{kill}}

\newcommand{\cmark}{\ding{51}}%
\newcommand{\xmark}{\ding{55}}%
\newcommand{\mscore}{\mathcal{MS}}
\newcommand{\tests}{\mathcal{T}}
\newcommand{\test}{{t}}
\newcommand{\prog}{\mathcal{P}}
\newcommand{\locs}{\mathcal{L}}
\newcommand{\progs}{\mathbb{P}}
\newcommand{\patch}{\mathcal{P'}}
\newcommand{\type}{\tau}
\newcommand{\rankedProgs}{\bar{\mathbb{P}}}
\newcommand{\muts}{\mathcal{M}}
\newcommand{\fixes}{\progs_{\scalebox{.5}{\cmark}}}
\newcommand{\mut}{m}
\newcommand{\exec}[1]{\llbracket #1 \rrbracket}
\newcommand{\mutgen}{{\tt MutGen}}
\newcommand{\faultloc}{{\tt FaultLocalization}}
\newcommand{\diff}{{\tt Diff}}
\newcommand{\cover}{{\tt Cover}}
\newcommand{\failing}{{\tt failing}}
\newcommand{\passing}{{\tt passing}}
\newcommand{\falsified}{{falsified}}
\newcommand{\true}{{\tt True}}
\newcommand{\false}{{\tt False}}
\newcommand{\nothing}{\APLbox}
\newcommand{\mutate}{$\hookrightarrow$}
\newcommand{\infer}[3]{${#1}\vdash{#2\hookrightarrow #3}$}
\newcommand{\inmeth}[2]{{#1}~\md\cjtext{(}...\cjtext{)\{}...{#2}...\cjtext{\}}}
\newcommand{\premeth}[2]{{#1}~\md\cjtext{(}t_1~e_1,...,t_n~e_n\cjtext{)\{}{#2}\cjtext{\}}}

%mutators
\newcommand{\argProb}{ARGUMENT PROPAGATION\xspace}
\newcommand{\retVal}{RETURN VALUE\xspace}
\newcommand{\consCall}{CONSTRUCTOR CALL\xspace}
\newcommand{\incre}{INCREMENTS\xspace}
\newcommand{\inCons}{INLINE CONSTANTS\xspace}
\newcommand{\memVar}{MEMBER VARIABLE\xspace}
\newcommand{\swi}{SWITCH\xspace}
\newcommand{\methCall}{METHOD CALL\xspace}
\newcommand{\invNeg}{INVERT NEGATIVES\xspace}
\newcommand{\ariOp}{ARITHMETIC OPERATOR\xspace}
\newcommand{\cond}{CONDITIONAL\xspace}
\newcommand{\derefGua}{DEREFERENCE GUARD\xspace}
\newcommand{\methGua}{METHOD GUARD\xspace}
\newcommand{\prePostCond}{PRE/POST- CONDITION\xspace}
\newcommand{\fieldAcc}{FIELD NAME\xspace}
\newcommand{\methName}{METHOD NAME\xspace}
\newcommand{\argList}{ARGUMENT LIST\xspace}
\newcommand{\locVar}{LOCAL VARIABLE\xspace}
\newcommand{\accessor}{ACCESSOR\xspace}
\newcommand{\caseBreaker}{CASE BREAKER\xspace}

\newcommand{\argProbS}{AP\xspace}
\newcommand{\retValS}{RV\xspace}
\newcommand{\consCallS}{CC\xspace}
\newcommand{\increS}{IS\xspace}
\newcommand{\inConsS}{IC\xspace}
\newcommand{\memVarS}{MV\xspace}
\newcommand{\swiS}{SW\xspace}
\newcommand{\methCallS}{MC\xspace}
\newcommand{\invNegS}{IN\xspace}
\newcommand{\ariOpS}{AO\xspace}
\newcommand{\condS}{CO\xspace}
\newcommand{\derefGuaS}{DG\xspace}
\newcommand{\methGuaS}{MG\xspace}
\newcommand{\prePostCondS}{PC\xspace}
\newcommand{\fieldAccS}{FN\xspace}
\newcommand{\methNameS}{MN\xspace}
\newcommand{\argListS}{AL\xspace}
\newcommand{\locVarS}{LV\xspace}
\newcommand{\accessorS}{AM\xspace}
\newcommand{\caseBreakerS}{CB\xspace}

%alg
\SetKw{Continue}{continue}
\SetKw{Break}{break}

%edit distance
\newcommand{\COPY}{\textsf{CPY}\xspace}
\newcommand{\KILL}{\textsf{KILL}\xspace}
\newcommand{\CHANGE}[1]{\textsf{CHG(#1)}\xspace}
\newcommand{\INSERT}[1]{\textsf{INS(#1)}\xspace}
\newcommand{\DELETE}{\textsf{DEL}\xspace}

%subs
\newcommand{\Chart}{\textsf{Chart}\xspace}
\newcommand{\Math}{\textsf{Math}\xspace}
\newcommand{\Lang}{\textsf{Lang}\xspace}
\newcommand{\Time}{\textsf{Time}\xspace}
\newcommand{\Closure}{\textsf{Closure}\xspace}
\newcommand{\Mockito}{\textsf{Mockito}\xspace}

\newcommand{\RQOne}{\textbf{RQ1}}
\newcommand{\RQTwo}{\textbf{RQ2}}
\newcommand{\RQThree}{\textbf{RQ3}}
\newcommand{\RQFour}{\textbf{RQ4}}

\newcommand{\ruledef}[2]{
            \fontsize{8pt}{8pt}\selectfont
            \cfrac{#1}{#2}
    }

\newcommand{\classicjava}{ClassicJava\xspace}
\newcommand{\cjtext}[1]{\textsf{\textbf{#1}}}

\begin{abstract}
Software debugging is tedious, time-consuming, and even error-prone by
itself. So, various automated debugging techniques have been proposed
in the literature to facilitate the debugging process. Automated
Program Repair (\apr) is one of the most recent advances in automated
debugging, and can directly produce patches for buggy programs with
minimal human intervention. Although various advanced \apr techniques
(including those that are either search-based or semantic-based) have
been proposed, the simplistic mutation-based \apr technique, which
simply uses pre-defined mutation operators (e.g., changing {\tt a>=b}
into {\tt a>b}) to mutate programs for finding patches, has not yet
been thoroughly studied. In this paper, we implement the first
practical bytecode-level \apr technique, \simpr, and present the first
extensive study on fixing real-world bugs (e.g., \defectsj{} bugs)
using bytecode mutation. The experimental results show that
surprisingly even \simpr with only the basic traditional mutators
can produce genuine patches for \gpatch bugs. Furthermore, with our
augmented mutators, \simpr is able to produce genuine patches for
\allgenuines bugs, significantly outperforming state-of-the-art \apr.
It is also an order of magnitude faster, indicating a promising
future for bytecode-mutation-based \apr.

%pro1: complete search space study
%pro2: bytecode level efficient \apr
%pro3: surprising number of fixed bugs (genuine fixes and plausible fixes with debugging hints)
\end{abstract}

\maketitle

\section{Introduction}\label{sec:Introduction}
%background about automated debugging
Software systems are ubiquitous in today's world; most of our
activities, and sometimes even our lives, depend on computers
controlled by the software. Unfortunately, software systems are
not perfect and often come with bugs. Software debugging is a
difficult activity that consumes over 50\% of the development time
and effort \cite{bib:WhitePaper}, and it costs the global economy
billions of dollars \cite{bib:Cambridge}. To date, a huge body of
research has been dedicated to automated debugging to automatically
localize \cite{bib:WGLW16,bib:AHMPP08,bib:XZ17,abreu2007accuracy,b2016learning,
li2016iterative,zhang2005experimental,sohn2017fluccs,zhang2017boosting}
or fix \cite{bib:TKKX14,bib:Weim06,bib:LDR14,bib:PKLABCPSSSWZER09,bib:GNFW12,
bib:DW10,bib:JCMY16,bib:TR15,bib:WCWHC18,bib:LR15,bib:LR16b,bib:NQRC13,
bib:XMDCMDBM17,bib:LPF17,bib:MYR16,bib:PFNWMZ14,bib:DZM09,bib:GMK11,
bib:GMM17,bib:Monp18,bib:MDSXM17} software bugs. Among various automated
debugging techniques, Automated Program Repair (\apr) techniques aim to
directly fix software bugs with minimal human intervention. These techniques can
significantly speed up the debugging process by either synthesizing
\emph{genuine} patches (i.e., the patches semantically equivalent to
developer patches) or suggesting patches, that may guide the debuggers
to fix the bugs faster. Thus,\Comment{ due to its benefits,} \apr has
been the subject of intense research in spite of being a young research
area.\Comment{Therefore, despite a young research area, \apr has been the
subject of intensive research.}

%existing repair techniques
Based on the actions they take for fixing a bug, state-of-the-art \apr
techniques can be divided into two broad categories: (1) techniques
that monitor the dynamic execution of a program to find deviations
from certain specifications, and then \emph{heal} the program by modifying its
runtime state in case of any abnormal behavior \cite{bib:LDR14,bib:PKLABCPSSSWZER09};
(2) so-called \emph{generate-and-validate} techniques that modify the code
representation of the programs based on various rules/techniques, and
then use either test cases or some formal specification (such as code
contracts) as an oracle to validate each generated candidate patch, and
find \emph{plausible} patches (i.e., the patches that can pass all the
tests/checks) \cite{bib:GNFW12,bib:DW10,bib:JCMY16,bib:TR15,bib:WCWHC18,
bib:LR15,bib:LR16b,bib:NQRC13,bib:XMDCMDBM17,bib:LPF17,bib:MYR16,
bib:PFNWMZ14,bib:DZM09,bib:GMK11}. Among these, generate-and-validate
techniques, especially those that are based on test cases, have gained
popularity as testing is the prevalent method for detecting software bugs,
while very few software systems are based on rigorous, formal specifications.

%limitations of existing \apr techniques (we want the limitations that our tech can address here)
It is worth noting that, lately, multiple \apr research papers get
published in top-notch Software Engineering conferences and journals
each year, introducing various delicately designed and/or implemented
\apr techniques. With such state-of-the-art \apr techniques, more and
more real bugs can be fixed fully automatically, e.g., the most recent
\apr technique published in ICSE'18 \cite{bib:WCWHC18} has been
reported to produce genuine patches for 22 bugs of \defectsj (a set
of real-world Java programs widely used for evaluating \apr techniques
\cite{bib:JJE14}). Despite the success of recent \apr techniques, as
also highlighted in a recent survey paper \cite{bib:GMM17}, currently
we have a \emph{scattered} collection of findings and innovations with
no thorough evaluation of the techniques or some clear relationship
between them. Actually, it is even not clear how the existing simplistic
mutation-based \apr \cite{bib:DW10}, which generates program patches simply
based on a limited set of mutators (also called mutation operators)
\cite{bib:AO08}, compares to the state-of-the-art \apr techniques.

%our basic motivation and idea
Therefore, in this work, we present the first extensive study of a
simplistic program repair approach that applies mutation-like patch
generation on the widely used \defectsj benchmark programs. More
specifically, we build a practical \apr tool named \simpr ({\bf
  Pra}ctical {\bf P}rogram {\bf R}epair) based on JVM bytecode
mutation. We \Comment{acknowledge}stress that although simplistic, \simpr offers
various benefits over the state-of-the-art techniques.  First, \simpr is
the first bytecode-level \apr technique for Java and all the
generated patches can be directly validated without compilation, while
existing techniques
\cite{bib:GNFW12,bib:DW10,bib:JCMY16,bib:TR15,bib:WCWHC18,
  bib:LR15,bib:LR16b,bib:NQRC13,bib:XMDCMDBM17,bib:LPF17,bib:MYR16,bib:PFNWMZ14,
  bib:DZM09,bib:GMK11} have to compile each candidate patch before
validating it\footnote{Please note that compilation can be quite
  expensive especially for type-safe programming languages.}.\Comment{ For
example,} Even though some techniques curtail compilation overhead by
encoding a group of patches inside a single meta-program, it can still
take up to 37 hours to fix a \defectsj program \emph{due to numerous
  patch compilations and class loadings} \cite{bib:LPF17}.  Second,
bytecode-level repair avoids messing up the source code in unexpected
ways, and can even be applicable for fixing code without source code
information, e.g., buggy 3rd-party libraries that do not have official
patches yet. Third, manipulating programs at the level of JVM bytecode
\cite{bib:LYBB17} makes \simpr independent of the syntax of a specific
target programming language, and makes it applicable to fix programs
written in other JVM-based languages (notably Kotlin
\cite{bib:Kotlin18}, Scala \cite{bib:Scala18}, and even Groovy
\cite{bib:Groovy18}). Lastly, \simpr does not require complex patching
rules \cite{bib:GNFW12,bib:LR15,bib:XMDCMDBM17}, complicated
computations such as symbolic execution and equation solving
\cite{bib:NQRC13,bib:MYR16,bib:LPF17}, or any training/mining
\cite{xiong2017precise,saha2017elixir, bib:WCWHC18}, making it
directly applicable for real-world programs and easily adoptable as
the baseline for future \apr techniques.

%our detailed study
We have applied \simpr to fix all the \djbug bugs available in
\defectsj.  Surprisingly, even the basic traditional mutators can already
produce plausible and genuine patches for \ppatch and \gpatch bugs,
respectively. With both the traditional and our augmented simple
mutators (e.g., replacing field accesses or method invocations),
\simpr successfully produces genuine patches for \allgenuines
\defectsj bugs, thereby significantly outperforming the state-of-the-art
\apr techniques (e.g., the most recent \capGen{} technique published
in ICSE'18 \cite{bib:WCWHC18} can only fix 22 bugs). In addition,
thanks to the bytecode-level manipulation, \simpr with only
single-thread execution is already 26.1X faster than the state-of-the-art
\capGen{} in terms of per-patch time cost. Even compared with the
recent technique \jaid{} \cite{bib:LPF17} that reduces compilation
overhead via grouping patches in meta-programs, \simpr is still
an order of magnitude faster (i.e., 15.7X).

\Comment{takes only 188 minutes on a commodity
  computer, with a reasonable memory footprint (less than 16GiB), to
  validate 107270 patches of the largest \defectsj subject program
  (i.e., the Google \Closure Compiler \cite{bib:Closure18}), while the
  recent advanced \jaid{} technique requires more that 1601.7 minutes
  to explore only 14464 candidate patches for the same bug on the same
  hardware platform.}

In summary, this paper makes the following contributions:
\begin{itemize}
\item {\bf Implementation.} We implement a simplistic, yet practically effective, program
  repair technique via bytecode mutation, \simpr, for JVM-based
  programming languages.
\item {\bf Study.} We perform the first extensive study of our \simpr technique on all the \djbug real bugs from the \defectsj benchmark.
\item {\bf Results.} Our study demonstrates that \simpr with only
  basic traditional mutators can successfully fix \gpatch bugs from
  \defectsj, while using our augmented mutators it can significantly
  outperform the state-of-the-art techniques---fixing \allgenuines
  bugs while being an order of magnitude faster. Our study also shows
  that even the non-genuine plausible fixes can help with manual
  debugging by giving high-quality fixing hints (e.g. precisely localizing
  the bug or even partially fixing it) to the debuggers.
\item {\bf Guidelines.} Our findings demonstrate for the first time
  that simple idea of mutation-based program repair can greatly
  complement the state-of-the-art \apr techniques in at least three
  aspects (namely effectiveness, efficiency, and applicability), and can
  inspire more work to advance \apr in this direction.
\end{itemize}

%what is program repair and why this is import
%what's the current state-of-the-art techs and trends for program repair
%what's the limitation of the existing work, complicated, time consuming (source-code level), no extensive study about search space (only Fan Long's work)
%all the above motivates our work on studying the strength of mutation for repair
%details about our study design and SURPRISING findings
%contribution list

\section{Related Work}\label{sec:Background}
In this section, we introduce the background on mutation testing
(\S \ref{sec:mut}) and generate-and-validate automated program
repair (\S \ref{sec:apr}).

\subsection{Mutation Testing}\label{sec:mut}
\emph{Mutation testing} \cite{bib:AO08} is a powerful method for
assessing the quality of a given test suite in detecting potential
software bugs. Mutation testing measures test suite quality via
injecting ``artificial bugs'' into the subject programs through mutating it.
The basic intuition is that the more artificial bugs that a test suite can detect,
the more likely is it to detect potential real bugs, hence the test suite
is of higher quality \cite{bib:JJIEHF14, bib:ABL05}.\Comment{
a test suite that can detect more artificial
bugs, it is more likely to detect more potential real bugs.} Central to mutation testing is the notion of
\emph{mutation operator}, aka \emph{mutator}, which is used to
generate artificial bugs to mimic real bugs. Applying a
mutator on a program results in a \emph{mutant} (or \emph{mutation}) of
the program---a variant of the program that differs from the original
program only in the injected artificial bug, e.g., replacing
\texttt{a+b} with \texttt{a-b} in one statement. In first-order mutation
testing \cite{bib:JM09}, the mutators can be treated as program transformers
that change the meaning of the input program by making small, single-pointed
changes to it. This implies that the resulting mutants should
also be syntactically valid, and also typeable, for it has to simulate
a valid buggy version of the program under test. This also indicates
that the mutators are highly dependent on the target programming
language. Therefore, given an original program $\prog$, mutation
testing will generate a set of program variants, $\muts$, where each
element is a valid mutant $\mut$.

Given the original program $\prog$, and a mutant $\mut\in\muts$ of the
program, a test suite $\tests$ is said to \emph{kill} mutant $\mut$ if
and only if there exists at least one test case $\test\in \tests$ such
that the observable final state of $\prog$ on $\test$ differs from
that of $\mut$ on $\test$, i.e., $\prog\exec{\test}\neq\mut\exec{\test}$.
Similarly, a mutant is said to \emph{survive} if no test case in $\tests$
can kill it. It might be the case that some of the survived mutants are
(semantically) \emph{equivalent} to the original program, hence the name
\emph{equivalent mutants}. Apparently, no test case can ever kill
equivalent mutants. By having the number of killed and equivalent
mutants for a given test suite $\tests$, one may easily compute a
\emph{mutation score} to evaluate the quality of $\tests$, i.e., the ratio
of killed mutants to all non-equivalent mutants
($\mscore=\frac{|\muts_{killed}|}{|\muts|-|\muts_{equivalent}|}$). Besides
its original application in test suite evaluation, recently mutation
testing has also been widely applied in various other areas, such as
simulating real bugs for software-testing experiments \cite{bib:JJIEHF14,
bib:ABL05}, automated test generation \cite{zhang2010test,papadakis2010automatic},
fault localization \cite{bib:XZ17, moon2014ask, papadakis2012using},
and even automated program repair \cite{bib:DW10}. When using mutation
testing for program repair, the input is a buggy program $\prog$ and its
corresponding test suite $\tests$ with failed tests due to the bug(s).
The output will be a specific mutant $\mut\in\muts$ that can pass all
the tests within $\tests$. Such resulting mutants can be potential patches
for the original buggy program $\prog$.

\subsection{Generate-and-Validate Program Repair}\label{sec:apr}
A software bug is a fault or flaw in a computer program that causes it to
produce an incorrect (unexpected) output. Automatic program repair (\apr)
\cite{bib:TKKX14,bib:Weim06,bib:LDR14,bib:PKLABCPSSSWZER09,bib:GNFW12,
bib:DW10,bib:JCMY16,bib:TR15,bib:WCWHC18,bib:LR15,bib:LR16b,bib:NQRC13,
bib:XMDCMDBM17,bib:LPF17,bib:MYR16,bib:PFNWMZ14,bib:DZM09,bib:GMK11,
bib:GMM17,bib:Monp18,bib:MDSXM17} is referred to a mostly automated
process that localizes such bug(s) inside a buggy program and edits
the program such that it produces correct (expected) outputs. Modern
generate-and-validate \apr techniques usually first utilize existing
fault localization \cite{abreu2007accuracy,bib:AHMPP08,bib:WGLW16}
techniques to identify the most suspicious program elements, and then
systematically change, insert, or delete various suspicious code elements
to search for a new program variant that can produce expected outputs.

In practice, test cases play a central role in both localizing
the bugs and also checking if a program variant behaves as expected---i.e.
test cases are also used as \emph{fix oracles}. Fault localization
techniques use the information obtained from both failing and passing
test cases to compute degrees of suspiciousness for each element of the
program. For example, so-called \emph{spectrum-based fault localization} techniques
\cite{bib:WGLW16}, that identify the program elements covered by more failed tests
and less passed tests as more suspicious, have been widely adopted by
various \apr techniques \cite{bib:MM16,bib:GMM17,bib:Monp18}. Modifying
a buggy program results in various \emph{candidate patches} that could be
verified by using the available test suite. A candidate patch that can pass
all the failing and passing tests within the original test suite is called
a \emph{plausible patch}, while a patch that not only passes the all tests in the
original test suite but is also semantically equivalent to the corresponding
programmer-written patch denotes a \emph{genuine patch} (or \emph{correct patch})
in the literature.

Note that, due to the so-called \apr \emph{overfitting} problem
\cite{bib:GMM17,bib:Monp18,bib:QLAR15}, not all plausible patches might be considered
genuine patches. Overfitting is a principal problem with the generate-and-validate \apr
techniques because of their dependence on the test suites to verify patches.
In practice, test suites are usually not perfect, in that a patch that passes
the all the test cases in the test suite may not generalize to other
potential tests of the program. Because of this problem, various advanced \apr
techniques \cite{bib:NQRC13,bib:LPF17,bib:MYR16,xin2017identifying} have
been proposed to mitigate the overfitting problem.

Based on different hypotheses, state-of-the-art generate-and-validate
\apr tools use a variety of techniques to generate or synthesize
patches. Some of them are based on the hypothesis that most bugs could
be solved by searching through all the potential candidate patches based on
certain patching rules \cite{bib:DW10,bib:GNFW12}, hence the name
\emph{search-based} \apr. Alternatively, \emph{semantic-based} techniques use deeper
semantical analyses (such as symbolic execution) to synthesize
conditions, or even more complex code snippets, that can pass all the test cases
\cite{bib:MYR16,bib:NQRC13,bib:XMDCMDBM17}. There are also various other
studies on \apr techniques: while some studies show that generating
patches just by deleting the original software functionality
\cite{bib:QLAR15,bib:QMLDZW14} can be effective, other studies
\cite{bib:GNFW12,bib:WCWHC18} demonstrate that fix ingredients
could be adopted from somewhere in the buggy program itself or even
some other programs based on the so-called \emph{plastic surgery}
hypothesis \cite{bib:BBDHS14}.

As discussed in \S \ref{sec:mut}, mutation testing has also been applied
for \apr \cite{bib:DW10}. The hypothesis for mutation-based APR is that
``\emph{if the mutators mimic programmer errors, mutating a defective
program can, therefore, fix it.}'' Several studies have been performed
on the mutation-based program repair \cite{bib:DW10,bib:MM16,bib:QMLDZW14} demonstrating
its feasibility. However, the existing studies
either concern mutation-based \apr on a set of small programs (e.g.,
from the Siemens Suite \cite{siemens}) with artificial bugs \cite{bib:DW10}
or apply only a limited set of mutators \cite{bib:MM16}. For example,
the most recent study \cite{bib:MM16} on mutation-based \apr with 3
mutators shows that it can only fix 4 (despite 17 plausible patches)
bugs of the widely used \defectsj benchmark. Furthermore, all the existing
studies \cite{bib:DW10,bib:MM16,bib:QMLDZW14} apply mutation testing at the source code level, which
can incur substantial compilation/class-loading overhead and is language-dependent.
In this work, we perform an extensive study on mutation-based \apr using
state-of-the-art mutators and efficient bytecode-level mutation. Working
at the level of JVM bytecode immediately obviates the need for compilation,
and enables us to explore much larger search spaces in reasonable time on
commodity hardware, thereby shedding light on the practicality of this simplistic
approach. Bytecode-level \apr is also beneficial in that it makes our tool
easily applicable to other popular JVM-based programming languages, such as
Kotlin, Groovy, and Scala.
%what's mutation testing, important concepts in mutation testing

\section{\simpr}\label{sec:simpr}
In this section, first we present the overall approach of \simpr
(\S \ref{sec:overall}), i.e. applying mutation testing to perform
automated program repair. Then, we discuss the design of the muators
(\S \ref{sec:mutator}), which make up the core of \simpr. Both our overall
approach and mutator design are simplistic (e.g., without any mining
or learning information) so that the readers can easily reproduce our
experimental results, and further build on top of \simpr.

\subsection{Overall Approach}\label{sec:overall}
\begin{algorithm}[t!]
\small
\caption{\label{alg:simpr} \simpr}
\KwIn{Original buggy program ${\prog}$, failing tests $\tests_f$, passing tests $\tests_p$\Comment{, plausible patch limit $n$} }
\KwOut{Plausible patch set $\fixes$}
\Begin{
$\locs\leftarrow \faultloc(\prog)$\tcp{Fault localization}
$\progs \leftarrow \mutgen(\prog, \locs)$   \tcp{Candidate patch generation}
\tcc{Perform validation for each candidate patch}
\For{$\patch\in \progs$}{
  \emph{\falsified}=\false \tcp{Whether the patch is falsified }
  $\tests'\leftarrow \cover(\diff(\patch,\prog))$\\
  \lIf{! $\tests'\supseteq \tests_f$}{
    \Continue
  } \tcc{Check if originally failed tests still fail}
  \For{$\test\in \tests_f$}{
    \If{$\patch\exec{\test}=\failing$}{
      \emph{\falsified}=\true\\
      \Break\tcp{Abort current patch validation}
    }
    }
    \lIf{\falsified=\true}{
    \Continue
    }
    \tcc{Check if any originally passed test fails}
     \For{$\test\in \tests_p\cap\tests'$}{
    \If{$\patch\exec{\test}=\failing$}{
      \emph{\falsified}=\true\\
      \Break\tcp{Abort current patch validation}
    }
    }
     \If{\falsified=\false}{
       $\fixes\leftarrow \patch$\tcp{Store the current plausible patch}
        }
  }

\Return{$\fixes$} \tcp{Return the resulting patch set}
}
\end{algorithm}
%\DrawBox{a}{b}

\Comment{\tcc{Terminate when $n$ plausible patches found}
       \lIf{$|\fixes|==n$}{
          \Break
        }}

The overall approach of \simpr is presented in Algorithm
\ref{alg:simpr}.  The algorithm inputs are the original buggy program
$\prog$ and its test suite $\tests$ that can detect the bug(s). For
the ease of illustration, we represent the passing and failing tests
in the test suite as $\tests_p$ and$\tests_f$, respectively. The
algorithm output is $\fixes$, a set of plausible patches that can pass
all the tests in $\tests$, and the developers can further inspect
$\fixes$ to check if there is any genuine patch. Shown in the
algorithm, Line 2 first computes and ranks the suspicious program
locations $\locs$ using off-the-shelf fault localization techniques
(e.g., Ochiai \cite{abreu2007accuracy} for this work). Line 3 then
exhaustively generates candidate patches $\progs$ for all suspicious
locations (i.e., the locations executed by any failed test) using our
mutators presented in \S \ref{sec:mutator}. Following prior \apr work
\cite{bib:WCWHC18,bib:LPF17,bib:MM16}, patches modifying the more
suspicious program locations obtain a higher rank. Then, Lines 4 to 18
iterate through each candidate patch to find potentially plausible
patches.

To ensure efficient patch validation, following prior work
\cite{bib:WCWHC18, bib:MM16}, each candidate patch is firstly executed
against the failed tests (Lines 8-11), and will only be executed
against the remaining tests once it passes all the originally failed
tests. The reason is that the originally failed tests are more likely
to fail again on candidate patches, whereas the patches failing any
test are already falsified, and do not need to be executed against the
remaining tests for sake of efficiency. Furthermore, we also apply two
additional optimizations that have already been widely used in the
mutation testing community (e.g., \pit \cite{bib:PIT18} and Javalanche
\cite{schuler2009javalanche}). First, all the candidate patches are
directly generated at the JVM bytecode level to avoid expensive
recompilation of a huge number of candidate patches. Second, \simpr
records the detailed coverage information for the patched locations
(i.e., statements in this work). Line 6 computes the tests covering
the patched location for each candidate
patch as $\tests'$.  Based on this information, a large number of test
executions can be reduced.  For failing tests, if $\tests'$ does not
subsume $\tests_f$, the candidate patch can be directly skipped since
the patched location is not covered by all failed tests and thus
cannot make all failed tests pass (Line 7); for passing tests, \simpr
only needs to check the patch against the tests covering the patched
location (Line 13) since the other passing tests do not touch the
patched location and will still pass. If the patch passes all the tests,
it will be recorded in the resulting plausible patch set $\fixes$.
Finally, \simpr returns $\fixes$ for manual inspection (Line 20).

Note that the bytecode-level patches include enough information for
the developers to confirm/reject the patches and apply them to the
source code. As shown in Figure \ref{fig:SimPRReport}, the two example
bytecode-level patches that \simpr provides (in the first half of the
figure) include sufficient debugging information, and it is trivial
for the developers to understand and apply the patches to the source
code. In addition, as shown in Figure \ref{fig:PRaPRFix}, \simpr{}
supports automatically decompiling the mutated bytecode to present
the patches in the context of the program text. The tool provides source code
line numbers as well so as to help the developers understand the
relationship between the decompiled and the original program text.

\begin{figure}
  \centering
  \begin{scriptsize}
  \begin{tabular}{l}
    \hline
    \texttt{PraPR (JDK 1.7) Fix Report - Sat Jun 23 08:25:46 CDT 2018}\\
    \texttt{Number of Plausible Fixes: 2}\\
    \texttt{Total Number of Patches: 1289}\\
    \texttt{================================================}\\
    %\texttt{In version \underline{10} of Lang 2 mutants, among 554 mutants,} \\
    %\texttt{might be considered patches:} \\
    \texttt{1. Mutator = METHOD CALL (\underline{removed call to}}\\
    \qquad\underline{\texttt{java/lang/Character::isWhitespace, and supplied default return value false}}), \\
    \texttt{File Name = \underline{org/apache/commons/lang3/time/FastDateParser.java},}\\
    \texttt{Line Number = \underline{307}.} \\
    \texttt{-----------} \\
    \texttt{2. Mutator = CONDITIONAL (\underline{removed conditional - replaced equality check}} \\
    \qquad\underline{\texttt{with false}}), \\
    \texttt{File Name = \underline{org/apache/commons/lang3/time/FastDateParser.java},}\\
    \texttt{Line Number = \underline{307}.}\\
    \hline
    \hline
    Contents of the file \texttt{org/apache/commons/lang3/time/FastDateParser.java}\\
    from source directory of \Lang 10: \\
    \\
    \texttt{305 for(int i= 0; i<value.length(); ++i) \{}\\
    \texttt{306    char c= value.charAt(i);}\\
    \texttt{307    if(Character.isWhitespace(c)) \{}\\
    \texttt{308 ...}
    \\
    \hline
  \end{tabular}
  \end{scriptsize}
  \caption{Two example patch reports automatically generated by \simpr (for the bug \Lang-10).
  Underlined parts convey sufficient information for locating and fixing the buggy
  \texttt{if}-statement shown in the bottom part of the figure.}\label{fig:SimPRReport}
\end{figure}

%mutation=org.pitest.mutationtest.engine.gregor.mutators.NonVoidMethodCallMutator(removed call to java/lang/Character::isWhitespace), mutatedMethod=org.apache.commons.lang3.time.FastDateParser::escapeRegex(Ljava/lang/StringBuilder;Ljava/lang/String;Z)Ljava/lang/StringBuilder;, lineNumber=307, fileName=FastDateParser.java 

%mutation=org.pitest.mutationtest.engine.gregor.mutators.RemoveConditionalMutator_EQUAL_ELSE(removed conditional - replaced equality check with false), mutatedMethod=org.apache.commons.lang3.time.FastDateParser::escapeRegex(Ljava/lang/StringBuilder;Ljava/lang/String;Z)Ljava/lang/StringBuilder;, lineNumber=307, fileName=FastDateParser.java 

\begin{figure}  
    \centering
    \begin{scriptsize}
        \begin{adjustbox}{width=0.5\textwidth}  
    \begin{tabular}{l}
      \hline
      \begin{lstlisting}[language=java]
appendQuoting(description);
*@\sout{description.appendText(wanted.toString());}@*
+++description.appendText(wanted == null ? "null" : wanted.toString());
appendQuoting(description);
      \end{lstlisting}\\  
      \hline
      \hline
      \begin{lstlisting}[language=java]
/*28*/ this.appendQuoting(description);
/*29*/ description.appendText(this.wanted == null?null:this.wanted.toString());
/*30*/ this.appendQuoting(description);
      \end{lstlisting}\\
      \hline
    \end{tabular}
\end{adjustbox}
    \end{scriptsize}
    \caption{Programmer-written fix for the bug \Mockito-29 (above the double line), and decompiled mutant
    corresponding to the fix generated by \simpr below it; source code line numbers are provided in the form
    of comments\Comment{ so as to help the programmer in locating the patch conveniently}}\label{fig:PRaPRFix}
  \end{figure}
  %mutation=org.pitest.mutationtest.engine.gregor.mutators.RemoveConditionalMutator_EQUAL_ELSE(removed conditional - replaced equality check with false), mutatedMethod=org.apache.commons.lang3.time.FastDateParser::escapeRegex(Ljava/lang/StringBuilder;Ljava/lang/String;Z)Ljava/lang/StringBuilder;, lineNumber=307, fileName=FastDateParser.java 

\subsection{\simpr Mutators}\label{sec:mutator}
All \simpr mutators aim to mutate the input Java programs via simple
transformation rules that only affect one program statement at a
time. All our mutators are implemented to work at the level of JVM
bytecode for sake of efficiency, and our implementation supports the
full set of JVM instructions and data types. For simplicity in
presentation though, we chose to present all our mutators in a core
Java language, named \classicjava \cite{bib:FKF98}. Our goal is to
describe the mutators using a minimal subset of Java so that the
functionality of the mutators could be described simply, yet
unambiguously. Figure \ref{fig:ClassicJavaSyntax} presents the
abstract syntax of \classicjava, extended with mutable variables,
\cjtext{return} command with its intuitive meaning, a number of
derived forms for conditionals (and also
\cjtext{switch}-\cjtext{case}), boolean expressions, arithmetic
expressions, and a \cjtext{fail} command that throws an uncaught
exception which causes the program to halt immediately. The full
definition of the operational semantics, as well as the type-rules and
an informal description of each of the language constructs, for the
core part of \classicjava, could be found in the original paper
\cite{bib:FKF98} introducing the language.

\begin{figure}[t!]
  \centering
  \begin{small}
    \begin{tabular}{rcl}
        $P$ & $=$ & ${\defn}^*$~$e$ \\
        $\defn$ & $=$ & \cjtext{class}~$c$~\cjtext{extends}~$c$~\cjtext{implements}~$i^*$\cjtext{\{}${\field}^*~{\meth}^*$\cjtext{\}}\\
        & & $\mid$~\cjtext{interface}~$i$~\cjtext{extends}~\cjtext{\{}${\meth}^*$\cjtext{\}}\\
        $\field$ & $=$ & $t~\fd$ \\
        $\meth$ & $=$ & $t~\md$\cjtext{(}${\argdec}^*$\cjtext{)\{}$\body$\cjtext{\}}\\
        $\argdec$ & $=$ & $t~\var$ \\
        $\body$ & $=$ & $e~\mid~$\cjtext{abstract}\\
        $e$ & $=$ & $\cst~\mid~\are~\mid~\be~\mid~$\cjtext{new}~$c~\mid~\var~\mid~e$\cjtext{.}$\fd~\mid~e$\cjtext{.}$\fd$\cjtext{=}$e$\\
        & & $\mid~e$\cjtext{.}$\md$\cjtext{(}$e^*$\cjtext{)}$~\mid~$\cjtext{super.}$md$\cjtext{(}$e^*$\cjtext{)}$~\mid~$\cjtext{let}~$\var$~\cjtext{=}~$e$~\cjtext{in}~$e$\\
        & & $\mid~$$~\be~$\cjtext{?}$~e~$\cjtext{:}$~e~\mid~$\cjtext{switch(}$e$\cjtext{)}~$($\cjtext{case}~$\cst$\cjtext{:}~$e)^*$~\cjtext{default:}~$e$ \\
        & & $\mid~$\cjtext{fail}$~\mid~$\cjtext{return}$~e~\mid~$\cjtext{if}$~\be~$\cjtext{then}$~e~$\cjtext{else}$~e~\mid~e$\cjtext{++}$~\mid~\dots$\\
        $\are$ & $=$ & $n~\mid~e~$\cjtext{+}$~e~\mid~$\cjtext{-}$e~\mid~e~$\cjtext{-}$~e~\mid~\dots$\\
        $\be$ & $=$ & \cjtext{!}$e~\mid~e~$\cjtext{\&\&}$~e~\mid~e~$\cjtext{=}$~e~\mid~e~$\cjtext{$\cjmath{<}$}$~e~\mid~\dots$\\
        $\var$ & $=$ & a variable name or \cjtext{this} \\
        $c$ & $=$ & a class name or \cjtext{Object} \\
        $i$ & $=$ & an interface name or \cjtext{Empty} \\
        $\fd$ & $=$ & a field name \\
        $\md$ & $=$ & a method name \\
        $t$ & $=$ & $c~\mid~i~\mid~$\cjtext{int}$~\mid~$\cjtext{boolean}\\
        $\cst$ & $=$ & n$~\mid~$\cjtext{true}$~\mid~$\cjtext{false}$~\mid~$\cjtext{null}\\
        $n$ & $=$ & an integer
    \end{tabular}
  \end{small}
  \caption{Abstract syntax of \classicjava}\label{fig:ClassicJavaSyntax}
\end{figure}

\begin{table*}
\begin{adjustbox}{width=1.0\textwidth}
  \begin{tabular}{lll}
    \hline
    ID& Mutator Name& Rules\\
    \hline\hline
 \argProbS&\argProb& \infer{\type(e_i)\preceq \type(e_0\cjtext{.}m\cjtext{(}e_1,\dots, e_n\cjtext{)}), 0\leq i\leq n}
  {e_0\cjtext{.}m\cjtext{(}e_1,\dots, e_n\cjtext{)}}{e_i}\\

\retValS&\retVal&\infer{\type(e)=\cjtext{boolean}\Comment{, e'\in\{\cjtext{true},\cjtext{false}, \cjtext{!}e, }}{\cjtext{return }e}{\cjtext{return } \cjtext{(}e~\cjtext{== false ? true : false}\cjtext{)}}\\
&&\infer{\type(e)=\cjtext{int}, e'\in\{\cjtext{0}, \cjtext{(}e\cjtext{)+1}, \cjtext{(}e\cjtext{ == 0 ? 1 : 0}\cjtext{)}\}}{\cjtext{return }e}{\cjtext{return }e'}\\
&&\infer{\type(e)=\cjtext{Object}, e'\in\{\cjtext{null}, \cjtext{(}e\cjtext{ == null ? fail : }e\cjtext{)}\}}{\cjtext{return }e}{\cjtext{return }e'}\\
\consCallS&\consCall&\infer{}{\cjtext{new }c\cjtext{()}}{\cjtext{null}}\\
\increS&\incre&\infer{\star,\star'\in\{\cjtext{++},\cjtext{-}\cjtext{-}\}, \star\neq \star', e\in\{var,var\star'\}}{var\star}{e}\\
&&\infer{\star,\star'\in\{\cjtext{++},\cjtext{-}\cjtext{-}\}, \star\neq \star', e\in\{var,\star' var\}}{\star var}{e}\\
\inConsS&\inCons&\infer{n'\in\{\cjtext{0}, \cjtext{1}, \cjtext{-1}, \cjtext{(}n~\cjtext{+~1)}, \cjtext{(}n~\cjtext{-~1)},\cjtext{(-}n\cjtext{)}\}}{n}{n'}\\
\memVarS&\memVar&\infer{\type(e_1.fd)=\cjtext{int}}{e_1.fd=e_2}{e_1.fd=\cjtext{0}}\\
&&\infer{\type(e_1.fd)=\cjtext{boolean}}{e_1.fd=e_2}{e_1.fd=\cjtext{false}}\\
&&\infer{\type(e_1.fd)\preceq\cjtext{Object}}{e_1.fd=e_2}{e_1.fd=\cjtext{null}}\\
\swiS&\swi&\infer{}{\cjtext{switch(}e\cjtext{)~case~}\cst_1\cjtext{:~}e_1~\dots~\cjtext{case}~\cst_n\cjtext{:}~e_n~\cjtext{default:}~e_d}{\cjtext{switch(}e\cjtext{)~case~}\cst_1\cjtext{:~}e_d~\dots~\cjtext{case}~\cst_n\cjtext{:}~e_d~\cjtext{default:}~e_1}\\
&&\infer{1\leq i\leq n}{\cjtext{switch(}e\cjtext{)~case~}\cst_1\cjtext{:~}e_1~\dots~\cjtext{case}~\cst_n\cjtext{:}~e_n~\cjtext{default:}~e_d}{\cjtext{switch(}e\cjtext{)~}\dots\cjtext{case}~\cst_i\cjtext{:}~e_d\dots~\cjtext{default:}~e_d}\\
\methCallS&\methCall&\infer{\type(e\cjtext{.}md\cjtext{(}e_1,\dots, e_n\cjtext{)})=\cjtext{boolean}}{e\cjtext{.}md\cjtext{(}e_1,\dots, e_n\cjtext{)}}{\cjtext{false}}\\
&&\infer{\type(e\cjtext{.}md\cjtext{(}e_1,\dots, e_n\cjtext{)})=\cjtext{int}}{e\cjtext{.}md\cjtext{(}e_1,\dots, e_n\cjtext{)}}{\cjtext{0}}\\
&&\infer{\type(e\cjtext{.}md\cjtext{(}e_1,\dots, e_n\cjtext{)})\preceq\cjtext{Object}}{e\cjtext{.}md\cjtext{(}e_1,\dots, e_n\cjtext{)}}{\cjtext{null}}\\
%&&\infer{\type(e\cjtext{.}md\cjtext{(}e_1,\dots, e_n\cjtext{)})=\cjtext{void}}{e\cjtext{.}md\cjtext{(}e_1,\dots, e_n\cjtext{)}}{\cjtext{\nothing}}\\
\invNegS&\invNeg&\infer{\type(\var)=\cjtext{int}}{\cjtext{-}var}{var}\\
\hline
\rowcolor{GrayOne}\ariOpS&\ariOp&\infer{\star, \star'\in\{\texttt{+},\texttt{-},\texttt{*},\texttt{/},\texttt{\%},\texttt{>>},\texttt{>>>},\texttt{<<},\texttt{\&},\texttt{|},\mathtt{\hat{~}}\}, \star\neq\star'}{e_1\star e_2}{e_1\star' e_2}\\
\rowcolor{GrayOne}&&\infer{\star\in\{\texttt{+},\texttt{-},\texttt{*},\texttt{/},\texttt{\%},\texttt{>>},\texttt{>>>},\texttt{<<},\texttt{\&},\texttt{|},\mathtt{\hat{~}}\}}{e_1\star e_2}{e_1}\\
\rowcolor{GrayOne}&&\infer{\star\in\{\texttt{+},\texttt{-},\texttt{*},\texttt{/},\texttt{\%},\texttt{>>},\texttt{>>>},\texttt{<<},\texttt{\&},\texttt{|},\mathtt{\hat{~}}\}}{e_1\star e_2}{e_2}\\
\rowcolor{GrayOne}\condS&\cond&\infer{e'\in\{\cjtext{true},\cjtext{false},!e\}, \type(e)=\cjtext{boolean}}{\cjtext{if(}e\cjtext{)}}{\cjtext{if(}e'\cjtext{)}}\\
\rowcolor{GrayOne}&&\infer{\star,\star'\in\{\leq,\geq,<,>,=,!=,\neq\},\star\neq \star'}{e_1\star e_2}{e_1\star'e_2}\\
\hline
\rowcolor{GrayTwo}\derefGuaS&\derefGua&\infer{\inmeth{t}{e.\fd},\defval(t)=v}{e.\fd}{\cjtext{(}e\cjtext{ == null ? return }v\cjtext{ : }e.\fd\cjtext{)}}\\
\rowcolor{GrayTwo}&&\infer{\inmeth{t}{e.\fd}, \type(\var)=t}{e.\fd}{\cjtext{(}e\cjtext{ == null ? return }\var\cjtext{ : }e.\fd\cjtext{)}}\\
\rowcolor{GrayTwo}&&\infer{\inmeth{t}{e.\fd_1}, \type(\cjtext{this.}\fd_2)=t}{e.\fd_1}{\cjtext{(}e\cjtext{ == null ? return this}.\fd_2\cjtext{ : }e.\fd_1\cjtext{)}}\\
\rowcolor{GrayTwo}&&\infer{\type(e.\fd)=t,\defval(t)=v}{e.\fd}{\cjtext{(}e\cjtext{ == null ? }v\cjtext{ : }e.\fd\cjtext{)}}\\
\rowcolor{GrayTwo}&&\infer{\type(e.\fd)=\type(\var)}{e.\fd}{\cjtext{(}e\cjtext{ == null ? }\var\cjtext{ : }e.\fd\cjtext{)}}\\
\rowcolor{GrayTwo}&&\infer{\type(e.\fd_1)=\type(\cjtext{this.}\fd_2)}{e.\fd_1}{\cjtext{(}e\cjtext{ == null ? this.}\fd_2\cjtext{ : }e.\fd_1\cjtext{)}}\\
\rowcolor{GrayTwo}\methGuaS&\methGua&\infer{\type(e\cjtext{.}md\cjtext{(}e_1,\dots, e_n\cjtext{)})=t,\defval(t)=v}{e\cjtext{.}md\cjtext{(}e_1,\dots, e_n\cjtext{)}}{\cjtext{(}e\cjtext{ == null ? return }v\cjtext{ : }e\cjtext{.}md\cjtext{(}e_1,\dots, e_n\cjtext{)}\cjtext{)}}\\
\rowcolor{GrayTwo}&&\infer{\type(e\cjtext{.}md\cjtext{(}e_1,\dots, e_n\cjtext{)})=\type(\var)}{e\cjtext{.}md\cjtext{(}e_1,\dots, e_n\cjtext{)}}{\cjtext{(}e\cjtext{ == null ? return }\var\cjtext{ : }e\cjtext{.}md\cjtext{(}e_1,\dots, e_n\cjtext{)}\cjtext{)}}\\
\rowcolor{GrayTwo}&&\infer{\type(e\cjtext{.}md\cjtext{(}e_1,\dots, e_n\cjtext{)})=\type(\cjtext{this.}\fd)}{e\cjtext{.}md\cjtext{(}e_1,\dots, e_n\cjtext{)}}{\cjtext{(}e\cjtext{ == null ? return this.}\fd\cjtext{ : }e\cjtext{.}md\cjtext{(}e_1,\dots, e_n\cjtext{)}\cjtext{)}}\\
\rowcolor{GrayTwo}&&\infer{\type(e\cjtext{.}md\cjtext{(}e_1,\dots, e_n\cjtext{)})=t,\defval(t)=v}{e\cjtext{.}md\cjtext{(}e_1,\dots, e_n\cjtext{)}}{\cjtext{(}e\cjtext{ == null ? }v\cjtext{ : }e\cjtext{.}md\cjtext{(}e_1,\dots, e_n\cjtext{)}\cjtext{)}}\\
\rowcolor{GrayTwo}&&\infer{\type(e\cjtext{.}md\cjtext{(}e_1,\dots, e_n\cjtext{)})=\type(\var)}{e\cjtext{.}md\cjtext{(}e_1,\dots, e_n\cjtext{)}}{\cjtext{(}e\cjtext{ == null ? }\var\cjtext{ : }e\cjtext{.}md\cjtext{(}e_1,\dots, e_n\cjtext{)}\cjtext{)}}\\
\rowcolor{GrayTwo}&&\infer{\type(e\cjtext{.}md\cjtext{(}e_1,\dots, e_n\cjtext{)})=\type(\cjtext{this.}\fd)}{e\cjtext{.}md\cjtext{(}e_1,\dots, e_n\cjtext{)}}{\cjtext{(}e\cjtext{ == null ? this.}\fd\cjtext{ : }e\cjtext{.}md\cjtext{(}e_1,\dots, e_n\cjtext{)}\cjtext{)}}\\
\rowcolor{GrayTwo}\prePostCondS&\prePostCond&$e'_1,...,e'_m\in\{e_i\mid t_i\preceq\cjtext{Object}\wedge 0\leq i\leq n\},\defval(t)=v$\\
\rowcolor{GrayTwo}&&\qquad\infer{}{\premeth{t}{e}}{\premeth{t}{\cjtext{(}e'_1\cjtext{ ==null || }...\cjtext{ || }e'_m\cjtext{==null) ? return }v \cjtext{ : }e}}\\
\rowcolor{GrayTwo}&&$\inmeth{t}{e\cjtext{.}md\cjtext{(}e_1,\dots, e_n\cjtext{)}},\type(e\cjtext{.}md\cjtext{(}e_1,\dots, e_n\cjtext{)})\preceq\cjtext{Object},\defval(t)=v$ \\
\rowcolor{GrayTwo}&&\qquad\infer{}{e\cjtext{.}md\cjtext{(}e_1,\dots, e_n\cjtext{)}}{\cjtext{(}e\cjtext{.}md\cjtext{(}e_1,\dots, e_n\cjtext{)==null ? return }v\cjtext{ : }e\cjtext{.}md\cjtext{(}e_1,\dots, e_n\cjtext{))}}\\
\rowcolor{GrayTwo}&&$\inmeth{t}{e\cjtext{.}md\cjtext{(}e_1,\dots, e_n\cjtext{)}},\type(e\cjtext{.}md\cjtext{(}e_1,\dots, e_n\cjtext{)})\preceq\cjtext{Object},\type(\var)=t$ \\
\rowcolor{GrayTwo}&&\qquad\infer{}{e\cjtext{.}md\cjtext{(}e_1,\dots, e_n\cjtext{)}}{\cjtext{(}e\cjtext{.}md\cjtext{(}e_1,\dots, e_n\cjtext{)==null ? return }\var\cjtext{ : }e\cjtext{.}md\cjtext{(}e_1,\dots, e_n\cjtext{))}}\\
\rowcolor{GrayTwo}&&$\inmeth{t}{e\cjtext{.}md\cjtext{(}e_1,\dots, e_n\cjtext{)}},\type(e\cjtext{.}md\cjtext{(}e_1,\dots, e_n\cjtext{)})\preceq\cjtext{Object},\type(\cjtext{this.}\fd)=t$ \\
\rowcolor{GrayTwo}&&\qquad\infer{}{e\cjtext{.}md\cjtext{(}e_1,\dots, e_n\cjtext{)}}{\cjtext{(}e\cjtext{.}md\cjtext{(}e_1,\dots, e_n\cjtext{)==null ? return this.}\fd\cjtext{ : }e\cjtext{.}md\cjtext{(}e_1,\dots, e_n\cjtext{))}}\\
\rowcolor{GrayTwo}\fieldAccS&\fieldAcc&\infer{fd_1\neq fd_2, \type(e\cjtext{.}fd_1)=\type(e\cjtext{.}fd_2)}{e\cjtext{.}fd_1}{e\cjtext{.}fd_2}\\
\rowcolor{GrayTwo}\methNameS&\methName&\infer{md\neq md', \type(md)=\type(md') }{e\cjtext{.}md\cjtext{(}e_1,\dots,e_n\cjtext{)}}{e\cjtext{.}md'\cjtext{(}e_1,\dots,e_n\cjtext{)}}\\
\rowcolor{GrayTwo}\argListS&\argList&$e_i'\in\{e_1,\dots,e_n\}\cup\{\var\mid\exists e_i.\type(var)=\type(e_i)\}\cup\{\cjtext{this.}\fd\mid\exists e_i.\type(\cjtext{this.}\fd)=\type(e_i)\}\cup\{\cjtext{0},\cjtext{false},\cjtext{null}\}$\\
\rowcolor{GrayTwo}&&\qquad\infer{}{e\cjtext{.}md\cjtext{(}e_1,\dots,e_n\cjtext{)}}{e\cjtext{.}md\cjtext{(}e_1',\dots,e_m'\cjtext{)}}\\
\rowcolor{GrayTwo}\locVarS&\locVar&\infer{var_1\neq var_2, \type(var_1)=\type(var_2)}{var_1}{var_2}\\
\rowcolor{GrayTwo}&&\infer{\type(\var)=\type(\cjtext{this.}fd)}{\var}{\cjtext{this.}fd}\\
\rowcolor{GrayTwo}\accessorS&\accessor&\infer{\type(e\cjtext{.}fd)=\type(e\cjtext{.}\md\cjtext{()})}{e\cjtext{.}fd}{e\cjtext{.}\md\cjtext{()}}\\
\rowcolor{GrayTwo}&&\infer{\type(e_2)=t,\type(\md)=t_r~\md(t)}{e_1\cjtext{.}fd\cjtext{=}e_2}{e_1\cjtext{.}\md\cjtext{(}e_2\cjtext{)}}\\
\rowcolor{GrayTwo}&&\infer{\type(e\cjtext{.}fd)=\type(\var)}{e\cjtext{.}fd}{\var}\\
\rowcolor{GrayTwo}\caseBreakerS&\caseBreaker&$\inmeth{t}{\cjtext{switch(}\dots\cjtext{)}~\dots~\cjtext{case}~\cst_i\cjtext{:}~e_i~\dots~\cjtext{default:}~e_d},\defval(t)=v$\\
\rowcolor{GrayTwo}&&\qquad\infer{}{\cjtext{case}~\cst_i\cjtext{:}~e_i}{\cjtext{case}~\cst_i\cjtext{: (let temp=}e_i\cjtext{ in return }v\cjtext{)}}\\
\rowcolor{GrayTwo}&&$\inmeth{t}{\cjtext{switch(}\dots\cjtext{)}~\dots~\cjtext{case}~\cst_i\cjtext{:}~e_i~\dots~\cjtext{default:}~e_d},\type(\var)=t$\\
\rowcolor{GrayTwo}&&\qquad\infer{}{\cjtext{case}~\cst_i\cjtext{:}~e_i}{\cjtext{case}~\cst_i\cjtext{: (let temp=}e_i\cjtext{ in return }\var\cjtext{)}}\\
\rowcolor{GrayTwo}&&$\inmeth{t}{\cjtext{switch(}\dots\cjtext{)}~\dots~\cjtext{case}~\cst_i\cjtext{:}~e_i~\dots~\cjtext{default:}~e_d},\type(\cjtext{this.}\fd)=t$\\
\rowcolor{GrayTwo}&&\qquad\infer{}{\cjtext{case}~\cst_i\cjtext{:}~e_i}{\cjtext{case}~\cst_i\cjtext{: (let temp=}e_i\cjtext{ in return this.}\fd\cjtext{)}}\\
\hline
  \end{tabular}
  \end{adjustbox}
\caption{\label{tab:mutators} Supported Mutators}
\end{table*}

Table \ref{tab:mutators} presents the details of the mutators
supported by \simpr. Since \simpr is built based on the
state-of-the-art mutation engine \pit \cite{bib:PIT18} for generating
and executing patches, \simpr mutators are implemented by directly
augmenting \pit mutators. Note that although a slightly different
categorization is used, the table includes all the official \pit
mutators. Note further that we do not present the mutators involving
some datatypes (e.g., $\cjtext{float}$ and $\cjtext{double}$) due to
the \classicjava syntax. In the table, the white block presents all
the mutators directly supported by \pit. The part highlighted with
light gray presents the mutators that are partially supported by \pit,
and are further augmented in \simpr to support more cases. For
example, although original \pit supports negating a conditional
statement, or changing $>$ into $\geq$ and vice versa, it does not
support changing a relational operator to any other ones
\cite{laurent2017assessing}, e.g., $>$ into $=$, which is specified in
traditional mutation testing literature
\cite{jia2011analysis}. Therefore, we augment \pit's original mutators
\textsf{MathMutator}, \textsf{ConditionalsBoundaryMutator}, and
\textsf{NegateConditionalsMutator} following a recent study on \pit
\cite{laurent2017assessing}. Finally, the dark gray part of the table
presents all the\Comment{ additional} mutators that are particular to
\simpr.\Comment{we implemented from scratch as they are not currently
  supported by \pit.}  Note that all the augmented \simpr{} mutators
are defined based on the syntax of \classicjava{} to handle potential bugs
at the expression level (since \simpr{} currently only supports single-edit patches).
Specifically, the new \derefGua, \methGua{}, and \prePostCond{} mutators add null-checks for
object instances to handle condition-related bugs, the new
\caseBreaker{} mutator handles \cjtext{switch}-related bugs, while the
other new mutators simply replace field accesses, local variables,
method invocations with other type-compatible ones within the
same class (or the mutated class) based on
\emph{plastic surgery hypothesis}~\cite{bib:BBDHS14}
for handling bugs by taking advantage of other expressions.

In Table \ref{tab:mutators}, each rule is represented in the form of
\infer{p}{e}{e'}, which denotes that when the premise $p$ holds, a candidate
patch can be generated via mutating a single instance of the expression $e$
to an expression $e'$ (note that all the other portions of the input
program remains unchanged). In the case of no premises, $p$ will be
omitted in the rule, e.g. as in \infer{}{e}{e'}. In addition, the polymorphic
operator $\type(\cdot)$ computes typing information if the input is an
expression and returns a type descriptor (i.e., the parameter types and
return types according to JVM specification \cite{bib:jvm}) when the
input is the fully qualified name of a method. The function $\defval(\cdot)$,
given a type-name, returns the default value corresponding to the type
as described in JVM specification \cite{bib:jvm}. $\type_1 \preceq \type_2$
denotes that type $\type_1$ is a subtype of $\type_2$. For example, the first
mutator specifies that when the type of parameter $e_i$ is a subtype of the
return type of $e_0\cjtext{.}m\cjtext{(}e_1,\dots,e_n\cjtext{)}$, the method
invocation can be directly mutated into its parameter $e_i$.

To make the definitions easier to follow, in Table \ref{tab:mexample}, we further
present concrete examples of some of the mutators. In what follows, we are
discussing design challenges and rationale for each of our mutators.

\parabf{\ariOp (\ariOpS)} \pit's original mutator \textsf{MathMutator}
simply replaces each arithmetic operator with another one (e.g.,
\texttt{+} is only mutated to \texttt{*}). We further augment it to mutate
every arithmetic operator with each of other compatible operators. Furthermore,
following a recent study \cite{laurent2017assessing}, we augment \ariOp to
implement Arithmetic Operation Deletion \cite{bib:AO08} (AOD) that deletes
each operand of a binary operator (e.g., \texttt{a+b} to \texttt{a} or
\texttt{b}). AOD is achieved by deleting the instruction corresponding to
the arithmetic operator, but before simply deleting the instruction we need to
prepare the JVM stack by keeping only one of the operands. This is achieved
by either popping the second operand off the stack or first swapping
the two operands and then popping the first operand off the stack.
We stress that popping and swapping are done differently based on the types of
the operands. For single word operands (such as \texttt{int}s, references, and
\texttt{float}s) we simply use \texttt{POP} and \texttt{SWAP} instructions, while
for double word operands (such as \texttt{double}s, and \texttt{long}s) we use
\texttt{POP2} and \texttt{DUP2\_X2; POP2} instructions, respectively.

\parabf{\cond (\condS)} A large part of this mutator is already implemented
by the existing mutators of the PIT, namely \textsf{NegateConditionals} and
\textsf{ConditionalsBoundary}. There are a few cases (e.g. replacing
\texttt{<} with \texttt{!=}) that are not addressed and prevent us
from achieving the full effect of a more general mutator, traditionally
known as Relation Operation Replacement \cite{bib:AO08}. This is achievable by
simply defining a helper mutator that does the missing replacements.

\begin{wrapfigure}{r}{0.13\textwidth}
  \begin{tabular}{l}
  \texttt{DUP}\\
  \texttt{IFNONNULL escape}\\
  $x$\texttt{LOAD }$n$\\
  $x$\texttt{RETURN}\\
  \texttt{escape:}\\
\end{tabular}
\end{wrapfigure}
\parabf{\derefGua (\derefGuaS)} This mutator mutates field dereference
sites so as to inject code checking if the base expression is
\cjtext{null} at a given site. The mutator is intended to prevent
\textsf{NullPointerException}s that are considered one of the most
common bugs in Java programs \cite{bib:HP04,bib:HSP05}. If it is
non-\cjtext{null} the injected code does nothing otherwise it does
either of the following: (1) returns the default value corresponding to
the return type of the mutated method; (2) returns the value of a
local variable visible at the mutation point whose type is
compatible with the return type of the mutated method; (3) returns the
value of a field whose type is compatible with the return type of the
mutated method; (4) uses the default value corresponding to the type
of the field being dereferenced instead of the field dereference
expression; (5) uses a local variable visible at the mutation point
whose type is compatible with that of the field being dereferenced
instead of the field dereference expression; (6) uses a field whose
type is compatible with that of the field being dereferenced instead
of the field dereference expression. The set of JVM instructions
shown in the figure right, illustrate the general form of the checking
code injected for case (2). The code replaces a \texttt{GETFIELD} instruction,
where $n$ is the index of a visible local variable to
be returned, while $x$, depending on the type of the field being
dereferenced, could be \texttt{I} (for \texttt{int}), \texttt{L} (for
\texttt{long}), and so on.\Comment{We stress that our implementation supports the
full set of JVM bytecode instructions.}

\parabf{\methGua (\methGuaS)} This mutator targets virtual method invocation
instructions. The mutator adds a check similar to that of \derefGuaS. Similar to
the mutator \derefGuaS, this mutator is intended to avoid \textsf{NullPointerException}s.

\parabf{\prePostCond (\prePostCondS)} As the name suggests, this mutator is
intended to add nullness checks for the object-typed parameters and what the method
returns, provided that it is a subtype of \cjtext{Object}, to avoid
\textsf{NullPointerException}s. The intuition is that \cjtext{null}
values are returned in cases of failure, and object-typed parameters are usually expected
to point to genuine objects rather than \cjtext{null} \cite{bib:Hoare09}.

\parabf{\fieldAcc (\fieldAccS)} This mutator mutates all field access
instructions, namely \texttt{GETFIELD}, \texttt{PUTFIELD}, \texttt{GETSTATIC},
and \texttt{PUTSTATIC}. Upon visiting a field access instruction, the mutator
loads the owner class of the field to extract all the information about its
fields. The mutator then selects a different visible field (e.g., \texttt{public}
fields), whose type is compatible with that of the current one. It is worth noting
that the newly selected field should be \texttt{static} if and only if the current field
is declared to be \texttt{static}. Finally, the field access instruction is mutated
such that it now accesses the new field. This mutator is intended to compensate
the programmer errors in selecting fields with similar names.

\parabf{\methName (\methNameS)} This mutator targets all kinds of method
invocation instructions (whether it is \texttt{static} or \texttt{virtual}).
The operational details and also the rationale of this mutator is similar
to \fieldAccS. Upon mutating a method invocation, the mutator selects another
method with a different name but with the same type descriptor to replace the
original one. It avoids nonsensical mutations; in particular, it does not
mutate constructor calls (as no other method with a different name can result
in a legal program), and also does not replace a call to a method with a call
to a constructor or a class initializer \cite{bib:jvm}.

\parabf{\argList (\argListS)} Unlike \methNameS, this mutator mutates
a call site to call another method with the same name, and compatible return type, but with different
parameter types (i.e., another overload of the callee). Similar to \methNameS, this
mutator is intended to mitigate programmer mistakes in choosing the right overload.
We take advantage of the utility library shipped with ASM bytecode manipulation
framework \cite{bib:ASM} to create temporary local variables so as to store the old
argument values. This mutator uses a variant of Levenshtein's edit distance algorithm
\cite{bib:Levenshtein} to find the minimal set of operations needed for reordering
these local variables or using some other value (such as the default value corresponding
to the type of a given parameter, a local variable visible at the call site, or a
field of appropriate type) in order to prepare the stack before calling newly selected
method.

\parabf{\locVar (\locVarS)} This mutator replaces the definition or use of a
variable with the definition or use of another (visible) variable, or a field,
with the same type. Obtaining the set of visible variables at the mutation point
is the most challenging part of implementing this mutator. We compute the set of
visible variables at each point of the method under mutation using a simple dataflow
analysis \cite{bib:Much97}, before doing the actual mutation.\Comment{ We describe this
analysis shortly.} We need this mutator for a reason similar to why we need
\fieldAccS. Note that sometimes we also need to replace the access to a local variable
with an access to a field. Modern IDEs such as Eclipse and IDEA do a good job in
distinguishing these two accesses by using different colors for each. But, in
general, this kind of mistakes are unavoidable.

\parabf{\accessor (\accessorS)} This mutator replaces a read access to
a field with either a call to a method (with no parameter and a return
type compatible with that of the accessed field) or a type-compatible
local variable visible at the point of mutation. Furthermore, it
replaces a write access to a field with a method that accepts an
argument with the type compatible with that of the field that has been
written on. Besides avoiding programmer mistakes in choosing between
field and local variable access, this mutator is principally intended
to avoid datarace bugs that are common in concurrent programs due to
incorrect synchronization between different threads
\cite{bib:vonPraun11}.

\parabf{\caseBreaker (\caseBreakerS)} This mutator is intended to mitigate
the unintended fall-through of control flow in \cjtext{switch}-\cjtext{case}
statements due to missing break or return statements. The mutator injects
appropriate return statement at the end of each \cjtext{case} clause.
\Comment{

  The mutators \methGua{},
  \derefGua{}, and \prePostCond add null-checks for object-typed expressions,
  while the others new mutators simply replace field accesses, local
  variables, method invocations with other type compatible ones within
  the same class based on the principles of the \emph{plastic surgery}
  hypothesis \cite{bib:BBDHS14}. Lastly, the mutator \caseBreaker inserts
  approprite return statements at the end of each \cjtext{case} clause so
  as to mitigate a common programming error due to unwanted fall through
    
We now discuss the implementation details of our augmented
mutators. These details will be mainly presented in stack-based Java bytecode
instructions so that the reader will know more about the operational details
of the proposed mutators:}

%\begin{wraptable}{r}{0.25\textwidth}
\begin{table}\center
%\small
\scalebox{1.0}{
\begin{tabular}{l|l}
\hline 
ID& Illustration\\
\hline\hline
\argProbS&\codeIn{y=obj.m(x)}\mutate \codeIn{y=x}\\
\retValS&\codeIn{return x}\mutate\codeIn{return x+1}\\
\consCallS&\codeIn{Object obj = new Object()}\mutate \codeIn{Object obj = null}\\
\increS&\codeIn{x++}\mutate \codeIn{x--}\\ 
\inConsS&\codeIn{int x=0} \mutate \codeIn{int x=1}\\
\memVarS&\codeIn{private int x = 1}\mutate \codeIn{private int x = 0}\\
\swiS&\codeIn{case 1: s1; default:s2;}\mutate\codeIn{case 1: s2; default:s1;}\\
\methCallS&\codeIn{int y=obj.m(x)}\mutate\codeIn{int y=0}\\
\invNegS&\codeIn{return -x}\mutate\codeIn{return x}\\
\hline
\ariOpS&\codeIn{z=x+y}\mutate \codeIn{z=x-y}\\
\condS&\codeIn{if(x>y)}\mutate \codeIn{if(x>=y)}\\
\hline
\derefGuaS&\codeIn{int x=obj.f}\mutate\codeIn{int x=(obj=null?0:obj.f)}\\
\methGuaS&\codeIn{int y=obj.m(x)}\mutate\codeIn{int y=(obj=null?0:obj.m(x))}\\
\prePostCondS&\codeIn{int m(T n)\{this.map.get(n)\}}\\
&\qquad\mutate\codeIn{int m(T n)\{n==null?return 0:this.map.get(n)\}}\\
\fieldAccS&\codeIn{int x=obj.f1}\mutate \codeIn{int x=obj.f2}\\
\methNameS&\codeIn{obj.m1(x)}\mutate \codeIn{obj.m2(x)}\\
\argListS&\codeIn{obj.m(x,y)}\mutate \codeIn{obj.m(x)}\\
\locVarS&\codeIn{int x=y}\mutate\codeIn{int x=z}\\
\accessorS&\codeIn{int x=obj.f}\mutate\codeIn{int x=obj.getF()}\\
\hline
\end{tabular}}
\caption{\label{tab:mexample}Mutator Illustration}
\end{table}
%\end{wraptable}
%Negate Conditionals Mutator&\codeIn{if(x<y)} \mutate\codeIn{if(x>=y)} \\
%Remove Conditional Mutator&\codeIn{if(x==y)}\mutate\codeIn{if(true)}\\
%Remove Increments Mutator&\codeIn{x++}\mutate{\tt \sout{x++}}\\
%Void Method Call Mutator&\codeIn{obj.m()}\mutate{\tt \sout{obj.m()}}\\

\Comment{
  \lingming{ LET's delete this}
  
  For this purpose, we have extended edit distance algorithm to operate on two
  arrays of parameter types. Edit operations for this algorithm are \COPY, \KILL,
  \CHANGE{$t$}, \INSERT{$t$}, and \DELETE, i.e., the \texttt{Copy}, \texttt{Kill},
  \texttt{Change}, \texttt{Insert} and \texttt{Delete} operations following the literature
  \cite{Thompson:1999:HCF:520533}.

\begin{wrapfigure}{r}{0.15\textwidth}
  \begin{tabular}{l}
  \texttt{//1 2 3 4}\\
  \texttt{DUP}
  \texttt{//1 2 3 4 4}\\
  \texttt{SWAP2}
  \texttt{//1 4 4 2 3}\\
  \texttt{SWAP}
  \texttt{//1 4 4 3 2}\\
  \texttt{SWAP2}
  \texttt{//1 3 2 4 4}\\
  \texttt{POP}
  \texttt{//1 3 2 4}
\end{tabular}
\end{wrapfigure}

\COPY corresponds to leaving the parameter on the stack, \KILL
corresponds to popping the remaining items off the stack after the
stack is prepared as expected by the new callee. \CHANGE{$t$} and
\INSERT{$t$}, where $t$ is an internal type name \cite{bib:LYBB17},
either push a default value corresponding to $t$ or brings a value
of type $t$ already in the stack to the desired position by swapping
values in the stack. Swapping values, in some cases, could be rather
challenging. For example, the right JVM code brings the third argument
to the place of second argument, when all arguments are of the same size. 

\begin{wrapfigure}{l}{0.15\textwidth}
  \begin{tabular}{l}
  \texttt{//1 2 3 4}\\
  \texttt{SWAP2}
  \texttt{//3 4 1 2}\\
  \texttt{SWAP}
  \texttt{//3 4 2 1}\\
  \texttt{POP}
  \texttt{//3 4 2}\\
  \texttt{DUP\_X2}
  \texttt{//2 3 4 2}\\
  \texttt{POP}
  \texttt{//2 3 4}
\end{tabular}
\end{wrapfigure}

The operation \DELETE deletes an argument at a specific
position. Implementing this operation could also be challenging in
certain cases. For example, in order to delete the first of four
arguments of the same size, we could do the operations on the left.
Note that our implementation has full support for different types of parameters.

}

\section{Experimental Setup}\label{sec:ExperimentalSetup}

\subsection{Research Questions}
\label{sec:RQs}

Our study investigates the following research questions:
\begin{enumerate}
    \item[\RQOne] How does \simpr perform in terms of effectiveness on automatically fixing real bugs? %effectiveness and efficiency; also report genuine fixes
    \item[\RQTwo] How does \simpr perform in terms of efficiency on automatically fixing real bugs?
    %\item[\RQTwo] How does \simpr rank genuine patches within its search space? %give some hint to the programmer; summarize 4 or 5 different case that mutation testing give hint--> directly at the patch location or somewhere deep in the call stack.
%  \item How effective is mutation-based program repair?
% practical aspects:
    \item[\RQThree] How does \simpr compare with state-of-the-art automated program repair techniques? % Which mutation operators are most effective in repair? Does enlarging search space improve effectiveness of the technique?
%    \item[\RQFour] How can we do even faster mutation based program repair? %prioritize statements based on e.g Ochiai and mutator type. E.g. stop testing mutants after first observed failed test case
\end{enumerate}

\subsection{Subject Systems}
We conduct our experimentation on the \defectsj \cite{bib:JJE14}
benchmark. \defectsj is a collection of six real-world programs
from GitHub with known, reproducible real bugs. These programs
are real-world projects developed over an extended period of
time, so they contain a variety of programming idioms and are a
good representative of those programs found randomly in the wild.
Therefore, \defectsj programs are suitable for evaluating the
effectiveness of candidate program repair techniques. Furthermore,
\defectsj programs are extensively used in peer-reviewed research
work~\cite{bib:WCWHC18,bib:LPF17,saha2017elixir,xiong2017precise,
le2016history,bib:MM16} for program repair thereby enabling us to
compare our results with those related work. Table
\ref{tab:Defects4J-Intro} lists the detailed statistics about the
\defectsj programs. In the table, Column ``ID'' presents the
identifiers to represent each benchmark program. Column ``Name''
presents the original names for the \defectsj programs. Column
``\#Bugs'' presents the number of bugs for each program. Finally,
Columns ``\#Tests'' and ``LoC'' present the number of tests
(i.e., JUnit test methods) and the lines of code for the {\tt HEAD}
buggy version (the most recent version) of each program.

\begin{table}
  \centering
  \begin{tabular}{|c||l|rrr|}
    \hline
    \textbf{ID} & \textbf{Name} & \textbf{\#Bugs}&\textbf{\#Tests}&\textbf{LoC} \\
    \hline\hline
    \Chart & JFreeChart & 26 &2,205&96K\\
    \Time & Joda-Time & 27 &4,130&28K\\
    \Mockito & Mockito framework & 38&1,366&23K \\
    \Lang & Apache Commons-Lang & 65 &2,245&22K\\
    \Math & Apache Commons-Math & 106 &3,602& 85K\\
    \Closure & Google Closure Compiler & 133 &7,927&90K\\
    \hline
    Total &  & 395 &21,475&344K\\
    \hline
  \end{tabular}
  \caption{Defects4J programs}\label{tab:Defects4J-Intro}
\end{table}

\subsection{Implementation}
\simpr has been implemented as a full-fledged program repair tool for
JVM bytecode. Currently it supports Maven-based \cite{bib:Maven18} Java and Kotlin
projects with JUnit, or TestNG \cite{bib:TestNG}, test suites.
Given any such program with at least one failed test, \simpr
can be applied using a single command ``{\tt mvn
  prapr-plugin:prapr}''. During the repair process, \simpr uses the
ASM bytecode manipulation framework \cite{bib:ASM} with Java Agent
\cite{bib:Agent} to collect coverage information, and supports the
Ochiai spectrum-based fault localization \cite{abreu2007accuracy}.
We have built \simpr \Comment{uses}on top of the state-of-the-art mutation testing engine \pit
\cite{bib:PIT18} because of the
following reasons: (1) \pit works directly at the level of JVM
bytecode, indicating that it directly mutates compiled bytecode rather
than source code and thus might save substantial compilation and class-loading time; (2) \pit
\Comment{uses Java Agent to} does perform both mutation generation and execution
on-the-fly; (3) \pit supports various optimizations, e.g.,
coverage-based test execution reduction and multithread mutant
execution; (4) \pit is the most robust and widely used mutation
testing tool both in academia and industry \cite{laurent2017assessing,
  bib:PIT18}.  Since \pit is originally implemented for assessing test
effectiveness through mutation testing, we needed to perform the
following optimizations, modifications and extensions to it so as to
integrate it with \simpr. First, we enabled \pit to mutate programs
with failed tests since the original \pit only accepts programs with
passing test suites (called {\em green} test suites in
\pit). Second, we enabled \pit to skip the mutants whose mutated
statements are not covered by failed tests (Line 7 in Algorithm
\ref{alg:simpr}), since they cannot change the behaviors of the failed
tests, and thus cannot make them pass. Third, we forced \pit to always
first execute the failed tests against a patch (Lines 8-16 in
Algorithm \ref{alg:simpr}), since we do not need to further validate a
patch if any failed test still fails. Fourth, we augmented the
original \pit mutators and implemented the additional new mutators
(Table \ref{tab:mutators}) using the ASM bytecode manipulation
framework\Comment{ \cite{bib:ASM}}.

All our experimentation is done on a Dell workstation with Intel Xeon CPU
E5-2697 v4@2.30GHz and 98GB RAM, running Ubuntu 16.04.4 LTS and Oracle
Java 64-Bit Server version 1.7.0\_80. It is also worth mentioning that
we run \simpr using both single thread and 4 threads \emph{exhaustively}
on all candidate patches to precisely measure the cost of running \simpr.

\section{Result Analysis}\label{sec:res}
\subsection{RQ1: \simpr Effectiveness} \label{sec:rq1}
%\begin{scriptsize}
\begin{table*}
\begin{adjustbox}{width=\textwidth}
\begin{tabular}{|l|l|rrrrr|rrrrr||l|l|rrrrr|rrrrr|}
%\begin{tabular}{|l|l||rrrrr||rrrrr|||l|l||rrrrr||rrrrr|}
\hline
&&\multicolumn{5}{c|}{Original Mutators}&\multicolumn{5}{c||}{All Mutators}&
&&\multicolumn{5}{c|}{Original Mutators}&\multicolumn{5}{c|}{All Mutators}\\
%&&\multicolumn{5}{c||}{Original Mutators}&\multicolumn{5}{c|||}{All Mutators}&
%&&\multicolumn{5}{c||}{Original Mutators}&\multicolumn{5}{c|}{All Mutators}\\
Sub.&BugID&Time&\#Patch&\#Plau.&\#Genu.&Mutator&Time&\#Patch&\#Plau.&\#Genu.&Mutator&
Sub.&BugID&Time&\#Patch&\#Plau.&\#Genu.&Mutator&Time&\#Patch&\#Plau.&\#Genu.&Mutator\\
\hline
\hline
\Chart & 1 & 110 & 866 & \cellcolor{GrayTwo}2 & \cellcolor{GrayTwo}1 & \cellcolor{GrayTwo}CO & 249 & 3555 & \cellcolor{GrayTwo}2 & \cellcolor{GrayTwo}1 & \cellcolor{GrayTwo}CO&\Closure & 130 & 2053 & 12885 & \cellcolor{GrayTwo}1 & \cellcolor{GrayTwo}1 & \cellcolor{GrayTwo}CO & 4752 & 42797 & \cellcolor{GrayTwo}7 & \cellcolor{GrayTwo}1 & \cellcolor{GrayTwo}CO \\
\Chart & 3 & 49 & 411 & 0 & 0 & & 71 & 1225 & \cellcolor{GrayTwo}1 & 0 &&\Closure & 133 & 730 & 4156 & 0 & 0 & & 1684 & 16225 & \cellcolor{GrayTwo}7 & 0 & \\
\Chart & 4 & 104 & 960 & 0 & 0 & & 212 & 3503 & \cellcolor{GrayTwo}5 & 0 & &\Lang & 6 & 95 & 158 & 0 & 0 & & 132 & 296 & \cellcolor{GrayTwo}1 & \cellcolor{GrayTwo}1 & \cellcolor{GrayTwo}LV \\
\Chart & 5 & 36 & 162 & \cellcolor{GrayTwo}2 & 0 & & 45 & 366 & \cellcolor{GrayTwo}3 & 0 &&\Lang & 7 & 52 & 745 & \cellcolor{GrayTwo}3 & 0 & & 98 & 1252 & \cellcolor{GrayTwo}6 & 0 & \\
\Chart & 7 & 43 & 416 & \cellcolor{GrayTwo}1 & 0 & & 65 & 1634 & \cellcolor{GrayTwo}17 & 0 &&\Lang & 10 & 60 & 497 & \cellcolor{GrayTwo}2 & \cellcolor{GrayTwo}2 & \cellcolor{GrayTwo}MC, CO & 99 & 1289 & \cellcolor{GrayTwo}2 & \cellcolor{GrayTwo}2 & \cellcolor{GrayTwo}MC, CO \\
\Chart & 8 & 35 & 163 & 0 & 0 & & 43 & 515 & \cellcolor{GrayTwo}4 & \cellcolor{GrayTwo}1 & \cellcolor{GrayTwo}AM&\Lang & 22 & 140 & 229 & \cellcolor{GrayTwo}2 & 0 & & 277 & 335 & \cellcolor{GrayTwo}2 & 0 & \\
\Chart & 11 & 33 & 93 & 0 & 0 & & 35 & 169 & \cellcolor{GrayTwo}2 & \cellcolor{GrayTwo}1 & \cellcolor{GrayTwo}LV&\Lang & 25 & 19 & 3 & 0 & 0 & & 20 & 21 & \cellcolor{GrayTwo}8 & 0 & \\
\Chart & 12 & 66 & 512 & \cellcolor{GrayTwo}1 & 0 & & 105 & 1910 & \cellcolor{GrayTwo}2 & \cellcolor{GrayTwo}1 & \cellcolor{GrayTwo}AM&\Lang & 26 & 33 & 631 & 0 & 0 & & 52 & 1482 & \cellcolor{GrayTwo}1 & \cellcolor{GrayTwo}1 & \cellcolor{GrayTwo}AL \\
\Chart & 13 & 51 & 626 & \cellcolor{GrayTwo}12 & 0 & & 88 & 2566 & \cellcolor{GrayTwo}16 & 0 &&\Lang & 27 & 34 & 672 & \cellcolor{GrayTwo}10 & 0 & & 76 & 1102 & \cellcolor{GrayTwo}13 & 0 & \\
\Chart & 15 & 185 & 2312 & \cellcolor{GrayTwo}1 & 0 & & 425 & 8210 & \cellcolor{GrayTwo}10 & 0 &&\Lang & 31 & 23 & 76 & \cellcolor{GrayTwo}1 & 0 & & 26 & 122 & \cellcolor{GrayTwo}1 & 0 & \\
\Chart & 17 & 40 & 319 & \cellcolor{GrayTwo}1 & 0 & & 54 & 831 & \cellcolor{GrayTwo}1 & 0 &&\Lang & 33 & 20 & 24 & 0 & 0 & & 20 & 31 & \cellcolor{GrayTwo}1 & \cellcolor{GrayTwo}1 & \cellcolor{GrayTwo}MG \\
\Chart & 24 & 31 & 43 & 0 & 0 & & 33 & 133 & \cellcolor{GrayTwo}2 & \cellcolor{GrayTwo}1 & \cellcolor{GrayTwo}LV&\Lang & 39 & 88 & 300 & \cellcolor{GrayTwo}5 & 0 & & 185 & 771 & \cellcolor{GrayTwo}13 & 0 & \\
\Chart & 25 & 480 & 7040 & \cellcolor{GrayTwo}141 & 0 & & 1299 & 24876 & \cellcolor{GrayTwo}245 & 0 &&\Lang & 43 & 2942 & 81 & \cellcolor{GrayTwo}2 & 0 & & 9873 & 288 & \cellcolor{GrayTwo}2 & 0 & \\
\Chart & 26 & 311 & 3368 & \cellcolor{GrayTwo}52 & 0 & & 705 & 11920 & \cellcolor{GrayTwo}102 & \cellcolor{GrayTwo}1 & \cellcolor{GrayTwo}MG&\Lang & 44 & 29 & 252 & \cellcolor{GrayTwo}2 & 0 & & 41 & 389 & \cellcolor{GrayTwo}4 & 0 & \\
\Closure & 1 & 1762 & 8495 & \cellcolor{GrayTwo}1 & 0 & & 3130 & 28153 & \cellcolor{GrayTwo}5 & 0 &&\Lang & 51 & 29 & 317 & \cellcolor{GrayTwo}4 & 0 & & 31 & 375 & \cellcolor{GrayTwo}5 & \cellcolor{GrayTwo}1 & \cellcolor{GrayTwo}CB \\
\Closure & 2 & 1634 & 11397 & 0 & 0 & & 3958 & 39711 & \cellcolor{GrayTwo}12 & 0 &&\Lang & 57 & 25 & 4 & 0 & 0 & & 25 & 10 & \cellcolor{GrayTwo}3 & \cellcolor{GrayTwo}1 & \cellcolor{GrayTwo}AM \\
\Closure & 3 & 2303 & 14609 & \cellcolor{GrayTwo}1 & 0 & & 5370 & 49228 & \cellcolor{GrayTwo}6 & 0 &&\Lang & 58 & 32 & 350 & \cellcolor{GrayTwo}2 & 0 & & 48 & 593 & \cellcolor{GrayTwo}2 & 0 & \\
\Closure & 5 & 1619 & 11094 & \cellcolor{GrayTwo}1 & 0 & & 3871 & 38296 & \cellcolor{GrayTwo}6 & 0 &&\Lang & 59 & 25 & 53 & 0 & 0 & & 27 & 137 & \cellcolor{GrayTwo}2 & \cellcolor{GrayTwo}1 & \cellcolor{GrayTwo}LV \\
\Closure & 7 & 632 & 3666 & \cellcolor{GrayTwo}2 & 0 & & 1229 & 15762 & \cellcolor{GrayTwo}5 & 0 &&\Lang & 60 & 33 & 261 & 0 & 0 & & 46 & 613 & \cellcolor{GrayTwo}1 & 0 & \\
\Closure & 8 & 1392 & 8396 & \cellcolor{GrayTwo}2 & 0 & & 3195 & 30353 & \cellcolor{GrayTwo}6 & 0 &&\Lang & 61 & 44 & 193 & 0 & 0 & & 68 & 481 & \cellcolor{GrayTwo}1 & 0 & \\
\Closure & 10 & 1337 & 9257 & \cellcolor{GrayTwo}5 & 0 & & 3085 & 31350 & \cellcolor{GrayTwo}6 & \cellcolor{GrayTwo}1 & \cellcolor{GrayTwo}MN&\Lang & 63 & 79 & 710 & \cellcolor{GrayTwo}14 & 0 & & 100 & 1390 & \cellcolor{GrayTwo}33 & 0 & \\
\Closure & 11 & 2339 & 15428 & \cellcolor{GrayTwo}6 & \cellcolor{GrayTwo}2 & \cellcolor{GrayTwo}MC, CO & 5931 & 52010 & \cellcolor{GrayTwo}11 & \cellcolor{GrayTwo}2 & \cellcolor{GrayTwo}MC, CO&\Math & 2 & 593 & 1201 & \cellcolor{GrayTwo}27 & 0 & & 647 & 2326 & \cellcolor{GrayTwo}27 & 0 & \\
\Closure & 12 & 2198 & 14122 & \cellcolor{GrayTwo}5 & 0 & & 5451 & 47710 & \cellcolor{GrayTwo}10 & 0 &&\Math & 5 & 1435 & 121 & 0 & 0 & & 1491 & 381 & \cellcolor{GrayTwo}3 & \cellcolor{GrayTwo}1 & \cellcolor{GrayTwo}FN \\
\Closure & 13 & 3626 & 26721 & \cellcolor{GrayTwo}12 & 0 & & 9472 & 84915 & \cellcolor{GrayTwo}16 & 0 &&\Math & 6 & 1436 & 153 & \cellcolor{GrayTwo}1 & 0 & & 1471 & 497 & \cellcolor{GrayTwo}2 & 0 & \\
\Closure & 14 & 506 & 2279 & 0 & 0 & & 776 & 8068 & \cellcolor{GrayTwo}1 & \cellcolor{GrayTwo}1 & \cellcolor{GrayTwo}FN&\Math & 8 & 1449 & 993 & \cellcolor{GrayTwo}7 & 0 & & 1491 & 1973 & \cellcolor{GrayTwo}9 & 0 & \\
\Closure & 15 & 2012 & 12405 & \cellcolor{GrayTwo}1 & 0 & & 4836 & 42112 & \cellcolor{GrayTwo}5 & 0 &&\Math & 18 & 1110 & 6485 & 0 & 0 & & 2309 & 17763 & \cellcolor{GrayTwo}6 & 0 & \\
\Closure & 17 & 2173 & 15957 & \cellcolor{GrayTwo}1 & 0 & & 5650 & 53970 & \cellcolor{GrayTwo}1 & 0 &&\Math & 20 & 1307 & 6104 & \cellcolor{GrayTwo}31 & 0 & & 2283 & 16596 & \cellcolor{GrayTwo}50 & 0 & \\
\Closure & 18 & 2251 & 14504 & \cellcolor{GrayTwo}2 & \cellcolor{GrayTwo}2 & \cellcolor{GrayTwo}CO & 5252 & 46270 & \cellcolor{GrayTwo}2 & \cellcolor{GrayTwo}2 & \cellcolor{GrayTwo}CO&\Math & 28 & 864 & 1670 & \cellcolor{GrayTwo}40 & 0 & & 1087 & 4617 & \cellcolor{GrayTwo}51 & 0 & \\
\Closure & 21 & 1442 & 9698 & \cellcolor{GrayTwo}16 & 0 & & 3517 & 33133 & \cellcolor{GrayTwo}21 & 0 &&\Math & 29 & 872 & 917 & \cellcolor{GrayTwo}3 & 0 & & 1301 & 2308 & \cellcolor{GrayTwo}4 & 0 & \\
\Closure & 22 & 1447 & 9646 & 0 & 0 & & 3512 & 33018 & \cellcolor{GrayTwo}1 & 0 &&\Math & 32 & 1003 & 7473 & \cellcolor{GrayTwo}7 & 0 & & 2543 & 25574 & \cellcolor{GrayTwo}11 & 0 & \\
\Closure & 29 & 1759 & 10429 & \cellcolor{GrayTwo}1 & 0 & & 3666 & 35168 & \cellcolor{GrayTwo}6 & 0 &&\Math & 33 & 780 & 1796 & 0 & 0 & & 922 & 5001 & \cellcolor{GrayTwo}1 & \cellcolor{GrayTwo}1 & \cellcolor{GrayTwo}AL \\
\Closure & 30 & 1900 & 11006 & \cellcolor{GrayTwo}1 & 0 & & 4188 & 37672 & \cellcolor{GrayTwo}5 & 0 &&\Math & 34 & 699 & 91 & 0 & 0 & & 702 & 240 & \cellcolor{GrayTwo}1 & \cellcolor{GrayTwo}1 & \cellcolor{GrayTwo}AM \\
\Closure & 31 & 1700 & 9575 & \cellcolor{GrayTwo}4 & \cellcolor{GrayTwo}1 & \cellcolor{GrayTwo}CO & 3972 & 30413 & \cellcolor{GrayTwo}9 & \cellcolor{GrayTwo}1 & \cellcolor{GrayTwo}CO&\Math & 39 & 291 & 2346 & \cellcolor{GrayTwo}2 & 0 & & 1217 & 6058 & \cellcolor{GrayTwo}15 & 0 & \\
\Closure & 33 & 2443 & 18014 & \cellcolor{GrayTwo}1 & 0 & & 6414 & 60314 & \cellcolor{GrayTwo}1 & 0 &&\Math & 40 & 299 & 878 & \cellcolor{GrayTwo}4 & 0 & & 389 & 2220 & \cellcolor{GrayTwo}7 & 0 & \\
\Closure & 35 & 2343 & 17148 & 0 & 0 & & 6081 & 58088 & \cellcolor{GrayTwo}1 & 0 &&\Math & 42 & 328 & 1622 & \cellcolor{GrayTwo}1 & 0 & & 442 & 4467 & \cellcolor{GrayTwo}2 & 0 & \\
\Closure & 36 & 4247 & 34486 & \cellcolor{GrayTwo}4 & 0 & & 11252 & 107270 & \cellcolor{GrayTwo}8 & 0 &&\Math & 49 & 265 & 713 & \cellcolor{GrayTwo}9 & 0 & & 308 & 1807 & \cellcolor{GrayTwo}15 & 0 & \\
\Closure & 38 & 539 & 3272 & \cellcolor{GrayTwo}1 & 0 & & 850 & 9998 & \cellcolor{GrayTwo}1 & 0 &&\Math & 50 & 257 & 372 & \cellcolor{GrayTwo}11 & \cellcolor{GrayTwo}1 & \cellcolor{GrayTwo}CO & 271 & 1098 & \cellcolor{GrayTwo}30 & \cellcolor{GrayTwo}1 & \cellcolor{GrayTwo}CO \\
\Closure & 40 & 1624 & 10438 & \cellcolor{GrayTwo}3 & 0 & & 3638 & 33819 & \cellcolor{GrayTwo}3 & 0 &&\Math & 52 & 211 & 1031 & \cellcolor{GrayTwo}1 & 0 & & 277 & 3343 & \cellcolor{GrayTwo}1 & 0 & \\
\Closure & 42 & 495 & 3466 & \cellcolor{GrayTwo}4 & 0 & & 966 & 13958 & \cellcolor{GrayTwo}4 & 0 &&\Math & 57 & 217 & 196 & \cellcolor{GrayTwo}2 & 0 & & 287 & 502 & \cellcolor{GrayTwo}2 & 0 & \\
\Closure & 45 & 1553 & 10806 & \cellcolor{GrayTwo}6 & 0 & & 3819 & 37053 & \cellcolor{GrayTwo}12 & 0 &&\Math & 58 & 672 & 3164 & \cellcolor{GrayTwo}1 & 0 & & 2994 & 9002 & \cellcolor{GrayTwo}5 & \cellcolor{GrayTwo}1 & \cellcolor{GrayTwo}AL \\
\Closure & 46 & 446 & 2608 & \cellcolor{GrayTwo}4 & \cellcolor{GrayTwo}2 & \cellcolor{GrayTwo}MC, CO & 809 & 11464 & \cellcolor{GrayTwo}6 & \cellcolor{GrayTwo}2 & \cellcolor{GrayTwo}MC, CO&\Math & 59 & 213 & 1949 & 0 & 0 & & 445 & 3370 & \cellcolor{GrayTwo}1 & \cellcolor{GrayTwo}1 & \cellcolor{GrayTwo}LV \\
\Closure & 48 & 2076 & 15429 & \cellcolor{GrayTwo}3 & 0 & & 5374 & 52122 & \cellcolor{GrayTwo}6 & 0 &&\Math & 62 & 69 & 900 & 0 & 0 & & 96 & 2581 & \cellcolor{GrayTwo}2 & 0 & \\
\Closure & 50 & 1271 & 7390 & \cellcolor{GrayTwo}1 & 0 & & 2728 & 26148 & \cellcolor{GrayTwo}6 & 0 &&\Math & 63 & 41 & 97 & \cellcolor{GrayTwo}10 & 0 & & 45 & 135 & \cellcolor{GrayTwo}23 & 0 & \\
\Closure & 59 & 3784 & 29058 & \cellcolor{GrayTwo}1 & 0 & & 9561 & 90792 & \cellcolor{GrayTwo}7 & 0 &&\Math & 64 & 150 & 2129 & 0 & 0 & & 439 & 6817 & \cellcolor{GrayTwo}3 & 0 & \\
\Closure & 62 & 109 & 130 & \cellcolor{GrayTwo}2 & \cellcolor{GrayTwo}1 & \cellcolor{GrayTwo}CO & 114 & 436 & \cellcolor{GrayTwo}2 & \cellcolor{GrayTwo}1 & \cellcolor{GrayTwo}CO&\Math & 65 & 144 & 1936 & \cellcolor{GrayTwo}1 & 0 & & 312 & 6098 & \cellcolor{GrayTwo}1 & 0 & \\
\Closure & 63 & 108 & 130 & \cellcolor{GrayTwo}2 & \cellcolor{GrayTwo}1 & \cellcolor{GrayTwo}CO & 111 & 436 & \cellcolor{GrayTwo}2 & \cellcolor{GrayTwo}1 & \cellcolor{GrayTwo}CO&\Math & 70 & 33 & 80 & 0 & 0 & & 34 & 223 & \cellcolor{GrayTwo}2 & \cellcolor{GrayTwo}1 & \cellcolor{GrayTwo}AL \\
\Closure & 64 & 2381 & 18746 & \cellcolor{GrayTwo}1 & 0 & & 5920 & 59197 & \cellcolor{GrayTwo}2 & 0 &&\Math & 71 & 502 & 1031 & \cellcolor{GrayTwo}6 & 0 & & 1542 & 3302 & \cellcolor{GrayTwo}22 & 0 & \\
\Closure & 66 & 1029 & 7556 & \cellcolor{GrayTwo}11 & 0 & & 2301 & 27478 & \cellcolor{GrayTwo}16 & 0 &&\Math & 73 & 33 & 444 & 0 & 0 & & 46 & 1174 & \cellcolor{GrayTwo}6 & 0 & \\
\Closure & 68 & 736 & 3287 & \cellcolor{GrayTwo}1 & 0 & & 1638 & 13427 & \cellcolor{GrayTwo}10 & 0 &&\Math & 74 & 904 & 4340 & \cellcolor{GrayTwo}2 & 0 & & 3642 & 12204 & \cellcolor{GrayTwo}3 & 0 & \\
\Closure & 70 & 1693 & 10388 & \cellcolor{GrayTwo}1 & \cellcolor{GrayTwo}1 & \cellcolor{GrayTwo}IC & 3561 & 35521 & \cellcolor{GrayTwo}1 & \cellcolor{GrayTwo}1 & \cellcolor{GrayTwo}IC&\Math & 75 & 29 & 189 & 0 & 0 & & 40 & 557 & \cellcolor{GrayTwo}1 & \cellcolor{GrayTwo}1 & \cellcolor{GrayTwo}MN \\
\Closure & 72 & 1289 & 11017 & \cellcolor{GrayTwo}1 & 0 & & 3261 & 35910 & \cellcolor{GrayTwo}6 & 0 &&\Math & 78 & 57 & 825 & \cellcolor{GrayTwo}8 & 0 & & 152 & 2658 & \cellcolor{GrayTwo}15 & 0 & \\
\Closure & 73 & 444 & 3093 & \cellcolor{GrayTwo}1 & \cellcolor{GrayTwo}1 & \cellcolor{GrayTwo}CO & 766 & 9365 & \cellcolor{GrayTwo}1 & \cellcolor{GrayTwo}1 & \cellcolor{GrayTwo}CO&\Math & 80 & 293 & 7410 & \cellcolor{GrayTwo}42 & 0 &  & 1318 & 18725 & \cellcolor{GrayTwo}59 & 0 &  \\
\Closure & 76 & 1272 & 8854 & \cellcolor{GrayTwo}2 & 0 & & 2991 & 30813 & \cellcolor{GrayTwo}8 & 0 &&\Math & 81 & 186 & 5707 & \cellcolor{GrayTwo}75 & 0 & & 941 & 14372 & \cellcolor{GrayTwo}187 & 0 & \\
\Closure & 81 & 397 & 2964 & \cellcolor{GrayTwo}4 & 0 & & 782 & 11789 & \cellcolor{GrayTwo}4 & 0 &&\Math & 82 & 60 & 1075 & \cellcolor{GrayTwo}3 & \cellcolor{GrayTwo}1 & \cellcolor{GrayTwo}CO & 108 & 2721 & \cellcolor{GrayTwo}7 & \cellcolor{GrayTwo}1 & \cellcolor{GrayTwo} CO \\
\Closure & 84 & 399 & 3003 & \cellcolor{GrayTwo}4 & 0 & & 794 & 12124 & \cellcolor{GrayTwo}4 & 0 &&\Math & 84 & 47 & 339 & \cellcolor{GrayTwo}4 & 0 & & 127 & 888 & \cellcolor{GrayTwo}4 & 0 & \\
\Closure & 86 & 435 & 1890 & \cellcolor{GrayTwo}3 & \cellcolor{GrayTwo}2 & \cellcolor{GrayTwo}IC, RV & 561 & 6332 & \cellcolor{GrayTwo}3 & \cellcolor{GrayTwo}2 & \cellcolor{GrayTwo}IC, RV&\Math & 85 & 150 & 702 & \cellcolor{GrayTwo}6 & \cellcolor{GrayTwo}1 & \cellcolor{GrayTwo}CO & 489 & 1599 & \cellcolor{GrayTwo}6 & \cellcolor{GrayTwo}1 & \cellcolor{GrayTwo}CO \\
\Closure & 92 & 905 & 7196 & \cellcolor{GrayTwo}3 & 0 & & 2451 & 24705 & \cellcolor{GrayTwo}4 & \cellcolor{GrayTwo}1 & \cellcolor{GrayTwo}MN&\Math & 88 & 69 & 1343 & \cellcolor{GrayTwo}1 & 0 & & 133 & 3217 & \cellcolor{GrayTwo}2 & 0 & \\
\Closure & 93 & 906 & 7196 & \cellcolor{GrayTwo}3 & 0 & & 2454 & 24706 & \cellcolor{GrayTwo}4 & \cellcolor{GrayTwo}1 & \cellcolor{GrayTwo}MN&\Math & 95 & 3027 & 841 & \cellcolor{GrayTwo}10 & 0 & & 24846 & 1614 & \cellcolor{GrayTwo}14 & 0 & \\
\Closure & 101 & 2129 & 16807 & \cellcolor{GrayTwo}2 & 0 & & 6290 & 53814 & \cellcolor{GrayTwo}6 & 0 &&\Math & 96 & 24 & 232 & \cellcolor{GrayTwo}1 & 0 & & 32 & 660 & \cellcolor{GrayTwo}1 & 0 & \\
\Closure & 107 & 2473 & 15207 & \cellcolor{GrayTwo}3 & 0 & & 5452 & 50516 & \cellcolor{GrayTwo}4 & 0 &&\Math & 101 & 21 & 155 & 0 & 0 & & 27 & 458 & \cellcolor{GrayTwo}3 & 0 & \\
\Closure & 108 & 2217 & 11317 & \cellcolor{GrayTwo}2 & 0 & & 5215 & 40139 & \cellcolor{GrayTwo}4 & 0 &&\Math & 104 & 67 & 546 & \cellcolor{GrayTwo}1 & 0 & & 260 & 1046 & \cellcolor{GrayTwo}1 & 0 & \\
\Closure & 109 & 863 & 3968 & 0 & 0 & & 1613 & 15702 & \cellcolor{GrayTwo}4 & 0 &&\Mockito & 8 & 36 & 171 & \cellcolor{GrayTwo}6 & 0 & & 54 & 458 & \cellcolor{GrayTwo}8 & 0 & \\
\Closure & 111 & 964 & 5688 & \cellcolor{GrayTwo}2 & 0 & & 1836 & 21900 & \cellcolor{GrayTwo}4 & 0 &&\Mockito & 15 & 93 & 1032 & \cellcolor{GrayTwo}1 & 0 & & 135 & 2910 & \cellcolor{GrayTwo}1 & 0 & \\
\Closure & 113 & 1224 & 7843 & \cellcolor{GrayTwo}2 & 0 & & 2743 & 28355 & \cellcolor{GrayTwo}2 & 0 &&\Mockito & 28 & 139 & 1374 & \cellcolor{GrayTwo}1 & 0 & & 200 & 3862 & \cellcolor{GrayTwo}1 & 0 & \\
\Closure & 115 & 2458 & 12164 & \cellcolor{GrayTwo}15 & 0 & & 4280 & 39721 & \cellcolor{GrayTwo}26 & 0 &&\Mockito & 29 & 129 & 1665 & 0 & 0 & & 181 & 4261 & \cellcolor{GrayTwo}6 & \cellcolor{GrayTwo}1 & \cellcolor{GrayTwo}MG \\
\Closure & 119 & 1714 & 9510 & \cellcolor{GrayTwo}3 & 0 & & 3621 & 33683 & \cellcolor{GrayTwo}3 & 0 &&\Mockito & 38 & 31 & 152 & 0 & 0 & & 40 & 464 & \cellcolor{GrayTwo}3 & \cellcolor{GrayTwo}1 & \cellcolor{GrayTwo}MG \\
\Closure & 120 & 1947 & 12532 & \cellcolor{GrayTwo}3 & 0 & & 4340 & 40877 & \cellcolor{GrayTwo}8 & 0 &&\Time & 4 & 86 & 1339 & \cellcolor{GrayTwo}6 & 0 & & 115 & 2619 & \cellcolor{GrayTwo}10 & \cellcolor{GrayTwo}1 & \cellcolor{GrayTwo}AL \\
\Closure & 121 & 1965 & 12532 & \cellcolor{GrayTwo}3 & 0 & & 4253 & 40877 & \cellcolor{GrayTwo}8 & 0 &&\Time & 11 & 138 & 3139 & \cellcolor{GrayTwo}44 & \cellcolor{GrayTwo}1 & \cellcolor{GrayTwo}MC & 175 & 5046 & \cellcolor{GrayTwo}56 & \cellcolor{GrayTwo}1 & \cellcolor{GrayTwo}MC \\
\Closure & 122 & 672 & 3570 & \cellcolor{GrayTwo}4 & 0 & & 1092 & 13945 & \cellcolor{GrayTwo}4 & 0 &&\Time & 14 & 73 & 1052 & 0 & 0 & & 88 & 1712 & \cellcolor{GrayTwo}1 & 0 & \\
\Closure & 123 & 727 & 5202 & 0 & 0 & & 1555 & 16573 & \cellcolor{GrayTwo}1 & 0 &&\Time & 17 & 205 & 4011 & \cellcolor{GrayTwo}6 & 0 & & 317 & 9835 & \cellcolor{GrayTwo}6 & 0 & \\
\Closure & 125 & 2868 & 19193 & \cellcolor{GrayTwo}1 & 0 & & 7025 & 63756 & \cellcolor{GrayTwo}3 & 0 &&\Time & 18 & 96 & 517 & \cellcolor{GrayTwo}3 & 0 & & 109 & 1119 & \cellcolor{GrayTwo}3 & 0 & \\
\Closure & 126 & 1635 & 8226 & \cellcolor{GrayTwo}8 & \cellcolor{GrayTwo}2 & \cellcolor{GrayTwo}MC, CO & 3025 & 28934 & \cellcolor{GrayTwo}16 & \cellcolor{GrayTwo}2 & \cellcolor{GrayTwo}MC, CO&\Time & 19 & 218 & 3312 & \cellcolor{GrayTwo}2 & \cellcolor{GrayTwo}1 & \cellcolor{GrayTwo}CO & 251 & 5740 & \cellcolor{GrayTwo}2 & \cellcolor{GrayTwo}1 & \cellcolor{GrayTwo}CO \\
\Closure & 127 & 1792 & 9142 & \cellcolor{GrayTwo}2 & 0 & & 3355 & 31768 & \cellcolor{GrayTwo}7 & 0 &&\Time & 20 & 308 & 4530 & 0 & 0 & & 435 & 10699 & \cellcolor{GrayTwo}3 & 0 & \\
\Closure & 129 & 3318 & 21923 & \cellcolor{GrayTwo}7 & 0 & & 7672 & 70265 & \cellcolor{GrayTwo}17 & 0 &&\Time & 24 & 176 & 2951 & \cellcolor{GrayTwo}4 & 0 & & 251 & 5304 & \cellcolor{GrayTwo}4 & 0 & \\
\hline
\cline{13-24}
\multicolumn{12}{c|}{}&\textbf{Overall} &  &  &  & 113 & 18 & &  &  & 148 & 43& \\
\cline{13-24}
\end{tabular}
\end{adjustbox}
\caption{\label{tab:main} Overall \simpr repair results}
\end{table*}

Table \ref{tab:main} presents the main repair results for all the bugs
that \simpr can generate plausible patches. In the table, Columns
``Sub.'' and ``BugID'' present the subject name and bug ID information
for each bug. Column ``Original Mutators'' presents the repair results
using only the original \pit mutators for each bug, including the
total repair time (using single thread) for validating all patches
(Column ``Time''), the number of all validated patches (Column
``\#Patch''), the number of plausible patches (Column ``\#Plau.''),
the number of genuine patches (Column ``\#Genu.''), and the mutators
that produce genuine patches (Column ``Mutator''). Note that we only
present the number of validated patches (i.e., the patches passing the check
at Line-7 in Algorithm \ref{alg:simpr}) in the paper\Comment{, which is much
smaller than the number of all candidate patches generated by \simpr},
since the other patches cannot pass all the failed tests and do not need to be validated. Similarly, the column
``All Mutators'' presents the corresponding repair results using all
the mutators (i.e., including both the original \pit and our augmented
mutators). Finally, the row ``Overall'' counts the total number of bugs for
which the original and all mutators can produce plausible or genuine
patches.

According to the table, surprisingly, even the original \pit mutators can
generate plausible patches for \ppatch bugs and genuine patches
for \gpatch bugs from the \defectsj benchmark, comparable to the most
recent work \capGen \cite{bib:WCWHC18} that produces genuine patches for 22 bugs.
On the contrary, prior work \cite{bib:MM16} showed that mutation
testing can only find 17 plausible patches and 4 genuine patches
for the \defectsj benchmark. One potential reason is that the prior
work was based on mutation at the level of source-code which might take much longer
time due to the expensive recompilation and class loading for each mutant, and thus does
not scale to large programs like \Closure. Another reason is that the prior
work used only a limited set of mutators (e.g., only three mutators
were considered), that cannot generate valid patches for many bugs
\footnote{A quick check indicates that had \jMut been able to scale to all
the \defectsj{} programs, it would generate up to 7 genuine fixes.}.
\emph{To our knowledge, this is the first study that demonstrates
the effectiveness of mutation testing for fixing real bugs.}

Furthermore, all the simple \simpr mutators (including both the
original \pit mutators and our augmented mutators) can produce
plausible and genuine patches for 148 and \allgenuines bugs,
respectively. To our knowledge, this is the largest number of bugs
reported as fixed for the \defectsj benchmark to date. We looked into
the reason and found that one main reason is \simpr's capability in
exploring such a large number of potential patches within a short time
due to the bytecode level patch generation/validation and our
execution optimizations. For example, for bug \Closure-70, \simpr with
a single thread is able to validate 35,521 candidate patches within an
hour (i.e., 10 patches per second!). \emph{This finding demonstrates
  the effectiveness of \simpr and shows the importance of fast (and
  \Comment{accurate}exhaustive) patch generation and validation for automatic program repair.}

We next show some example genuine patches, i.e. semantically
equivalent to the developer patches, produced by \simpr to qualitatively
illustrate the effectiveness of \simpr. As shown in Figure \ref{bug:t19},
\simpr using the mutator \cond is able to produce a genuine patch that is
identical to the patch provided by the developer. Also, \simpr using the mutator \retVal
produces a genuine patch that is semantically equivalent to
the actual developer patch, as shown in Figure \ref{bug:c86}. Note that
those patches are as expected for they directly fall into the
capability of the employed mutators. Interestingly, we also observe
various cases where \simpr is able to suggest complex genuine
patches. For example, Figure \ref{bug:c46} presents both the developer
and \simpr patches for \Closure-46. According to the figure, the developer
patch removes an overriding method from a subclass, which is hard to
directly model using \simpr mutators. Interestingly, the \simpr patch,
generated via the mutator \cond, is able to force the overriding method
to always directly invoke the corresponding overridden method in the
superclass---i.e. it is semantically equivalent to removing the overriding
method.

\begin{figure}
    \begin{lstlisting}[language=java]
         // Developer and PraPR patches
         *@\sout{\}\textbf{ else if} (offsetLocal > 0) \{}@*
      +++} else if (offsetLocal >= 0) {
          \end{lstlisting}
    \caption{\label{bug:t19} \Time-19 patches}
  \end{figure}

\begin{figure}
        \begin{lstlisting}[language=java]
         // Developer patch
         *@\sout{\textbf{return true};}@*
      +++return false;
         // PraPR patch
         *@\sout{\textbf{return true};}@*
      +++return true == false ? true : false;
        \end{lstlisting}
    \caption{\label{bug:c86}\Closure-86 patches}
  \end{figure}

\begin{figure}
  \begin{lstlisting}[language=java]
        // Developer patch
        *@\sout{\textbf{@Override}}@*
        *@\sout{\textbf{public} JSType getLeastSupertype(JSType that) \{}@*
            *@\sout{\textbf{if} (!that.isRecordType()) \{}@*
                *@\sout{\textbf{return super}.getLeastSupertype(that);}@*
            *@\sout{\}...\}}@*
        // PraPR patch
        @Override
        public JSType getLeastSupertype(JSType that) {
            *@\sout{\textbf{if} (!that.isRecordType()) \{}@*
        +++if (!false) {
                return super.getLeastSupertype(that);
            }...}
        \end{lstlisting}
  %    \end{scriptsize}}
  \caption{\label{bug:c46} \Closure-46 patches}
  \end{figure}

\begin{comment}
  \begin{figure}
        \begin{lstlisting}[language=java]
        // Developer patch
        *@\sout{\textbf{boolean} wasWhite= {\bf false};}@*...
        *@\sout{\textbf{if}(Character.isWhitespace(c)) \{...\}}@*
        *@\sout{wasWhite= {\bf false};}@*
        // PraPR patch
        boolean wasWhite= false;...
        *@\sout{\textbf{if}(Character.isWhitespace(c)) \{...\}}@*
     +++if(false) {...}
        wasWhite= false;
        \end{lstlisting}
    \caption{\label{bug:l10}\Lang-10 patches}
  \end{figure}
\end{comment}

\subsection{RQ2: \simpr Efficiency}\label{sec:rq2}
Since Table \ref{tab:main} only presents the results for the bugs with
plausible patches, in order to further study the efficiency of \simpr, in Table \ref{tab:time}, we also present
the efficiency information of the tool on all the \defectsj bugs.
To better understand the efficiency of \simpr, we apply
the tool both with a single thread (default setting) and 4 threads of execution.
In the table, the column ``Sub.'' presents the subject systems. The column
``Original Mutators'' presents the average number of all validated
patches (Column ``\#Patches'') and the average time cost when using 1
thread (Column ``1-T'') and 4 threads (Column ``4-T'') for all bugs of
each subject system using the original \pit mutators. Note that we also
include the detailed speedup of using 4 threads over using a single thread
in parentheses. Similarly, the column ``All Mutators'' presents the information
when using all \simpr mutators. According to the table, we find that \simpr is
remarkably efficient even using only a single thread. For example, \simpr with
all the mutators only takes less than 1 hour to validate all the 29,850 patches
for \Closure when using a single thread. In addition, using 4 threads can
further improve the efficiency of \simpr. For example, \Closure has
the 2.1X speedup when using 4 threads, since it has more patches and can
better utilize the concurrent patch validation. Note that besides the machine
execution time, the repair efficiency also involves the manual efforts in
inspecting the plausible patches. Thus, the ranking of the genuine patches
within all the validated/plausible patches is of particular importance to
truly understand the efficiency of \simpr. In the remainder of this section,
we discuss this aspect of efficiency.\Comment{efficiency, and is covered in the next paragraph.}

\begin{table}
\begin{adjustbox}{width=\columnwidth}
\begin{tabular}{|l||rrr||rrr|}
\hline
&\multicolumn{3}{c||}{Original Mutators}&\multicolumn{3}{c|}{All Mutators}\\
\cline{2-7}
Sub.&\#Patches&1-T&4-T &\#Patches&1-T&4-T \\
\hline
\hline
\Chart & 806.6 & 79.5 & 79.6(1.0X) & 2827.6 & 157.8 & 85.5(1.8X) \\
\Closure & 8828.4 & 1359.1 & 978.2(1.4X) & 29849.9 & 3027.3 & 1458.9(2.1X) \\
\Lang & 262.7 & 86.5 & 87.4(1.0X) & 544.4 & 210.2 & 198.8(1.1X) \\
\Math & 1296.5 & 933.1 & 611.1(1.5X) & 3333.2 & 1629.2 & 916.5(1.8X) \\
\Mockito & 968.5 & 106.2 & 57.3(1.9X) & 2601.0 & 148.8 & 77.4(1.9X) \\
\Time & 1563.0 & 112.6 & 67.1(1.7X) & 2968.2 & 144.2 & 87.3(1.7X) \\
\hline
\end{tabular}
\end{adjustbox}
\caption{\label{tab:time} Average \simpr time cost (s)}
\end{table}

\begin{table}
\begin{adjustbox}{width=0.85\columnwidth}
\begin{tabular}{|l|l||r|rr||r|rr|}
\hline
&&\multicolumn{3}{c||}{Original Mutators}&\multicolumn{3}{c|}{All Mutators}\\
\cline{3-8}
Sub.&BugID&\#Patch&\multicolumn{2}{c||}{Rank} &\#Patch&\multicolumn{2}{|c|}{Rank} \\
\hline
\hline
\Chart & 1 & 866 & 61 & (1) & 3555 & 306 & (1)\\
\Chart & 8 & 163 & N/A & (N/A) & 515 & 103 & (4)\\
\Chart & 11 & 93 & N/A & (N/A) & 169 & 168 & (2)\\
\Chart & 12 & 512 & N/A & (N/A) & 1910 & 174 & (2)\\
\Chart & 24 & 43 & N/A & (N/A) & 133 & 109 & (2)\\
\Chart & 26 & 3368 & N/A & (N/A) & 11920 & 1420 & (23)\\
\Closure & 10 & 9257 & N/A & (N/A) & 31350 & 2059 & (1)\\
\Closure & 11 & 15428 & 2808 & (4) & 52010 & 8777 & (5)\\
\Closure & 14 & 2279 & N/A & (N/A) & 8068 & 1 & (1)\\
\Closure & 18 & 14504 & 9450 & (1) & 46270 & 28358 & (1)\\
\Closure & 31 & 9575 & 4919 & (2) & 30413 & 20734 & (9)\\
\Closure & 46 & 2608 & 28 & (1) & 11464 & 89 & (1)\\
\Closure & 62 & 130 & 30 & (1) & 436 & 70 & (1)\\
\Closure & 63 & 130 & 30 & (1) & 436 & 70 & (1)\\
\Closure & 70 & 10388 & 292 & (1) & 35521 & 996 & (1)\\
\Closure & 73 & 3093 & 136 & (1) & 9365 & 242 & (1)\\
\Closure & 86 & 1890 & 1 & (1) & 6332 & 1 & (1)\\
\Closure & 92 & 7196 & N/A & (N/A) & 24705 & 429 & (2)\\
\Closure & 93 & 7196 & N/A & (N/A) & 24706 & 429 & (2)\\
\Closure & 126 & 8226 & 9 & (2) & 28934 & 50 & (5)\\
\Closure & 130 & 12885 & 5084 & (1) & 42797 & 16240 & (7)\\
\Lang & 6 & 158 & N/A & (N/A) & 296 & 220 & (1)\\
\Lang & 10 & 497 & 1 & (1) & 1289 & 1 & (1)\\
\Lang & 26 & 631 & N/A & (N/A) & 1482 & 1 & (1)\\
\Lang & 33 & 24 & N/A & (N/A) & 31 & 1 & (1)\\
\Lang & 51 & 317 & N/A & (N/A) & 375 & 375 & (5)\\
\Lang & 57 & 4 & N/A & (N/A) & 10 & 1 & (1)\\
\Lang & 59 & 53 & N/A & (N/A) & 137 & 112 & (2)\\
\Math & 5 & 121 & N/A & (N/A) & 381 & 116 & (1)\\
\Math & 33 & 1796 & N/A & (N/A) & 5001 & 795 & (1)\\
\Math & 34 & 91 & N/A & (N/A) & 240 & 35 & (1)\\
\Math & 50 & 372 & 40 & (11) & 1098 & 95 & (30)\\
\Math & 58 & 3164 & N/A & (N/A) & 9002 & 670 & (4)\\
\Math & 59 & 1949 & N/A & (N/A) & 3370 & 1 & (1)\\
\Math & 70 & 80 & N/A & (N/A) & 223 & 18 & (1)\\
\Math & 75 & 189 & N/A & (N/A) & 557 & 35 & (1)\\
\Math & 82 & 1075 & 393 & (3) & 997 & 742 & (7)\\
\Math & 85 & 702 & 324 & (6) & 1599 & 742 & (6)\\
\Mockito & 29 & 1665 & N/A & (N/A) & 4261 & 94 & (1)\\
\Mockito & 38 & 152 & N/A & (N/A) & 464 & 17 & (2)\\
\Time & 4 & 1339 & N/A & (N/A) & 2619 & 545 & (10)\\
\Time & 11 & 3139 & 31 & (1) & 5046 & 133 & (1)\\
\Time & 19 & 3312 & 1926 & (2) & 5740 & 3457 & (2)\\
\hline
\textbf{Avg.} & & 3038.6 & 1420.2 & (2.3) & 9696.5 & 2076.4 & (3.6)\\ \hline
\end{tabular}
\end{adjustbox}
\caption{\label{tab:rank} Rank of \simpr genuine fixes}
\end{table}

Table \ref{tab:rank} presents the ranking of the genuine patches among
all validated patches and all plausible patches. In the table, the columns
``Sub.'' and ``BugID'' present the bug version information. The column
``Original Mutators'' presents the number of validated patches (Column
``\#Patch'') and the rank of the first genuine patch among validated
patches (Column ``Rank'') when using the original \pit mutators. The
rank of the first genuine patch among all plausible patches is shown
in parentheses. Similarly, the column ``All Mutators'' presents the
corresponding results when using all \simpr mutators. Note that for
the case of tied patches, \simpr favors the patches generated by mutators
with smaller ratios of plausible to validated patches since the mutators
with larger ratios tend to be resilient to the corresponding test suite.
If the tie remains, \simpr uses the worst ranking for all tied patches.
From the table, we can observe that the genuine patches are ranked high
among validated and plausible patches when using both original and all
mutators. For example, on average, using all mutators, the genuine
patches are ranked 2076.4th among all the 9696.5 validated patches
(i.e., top 21.4\%) using all mutators. This observation demonstrates that
simply using Ochiai spectrum-based fault localization can provide an
effective ranking for the genuine patches, further confirming the
effectiveness of Ochiai for program repair. We also observe that the rank
of the genuine patches among all validated patches when using all mutators
are not significantly worse than when using only the original \pit
mutators. For example, among all validated patches, the genuine patches
are ranked 1420.2th using original mutators and 2076.4th using all
mutators on average. One potential reason is that using all mutators
provide more genuine patches, which are ranked high in the search
space. Furthermore, surprisingly, among the plausible patches, the
genuine patches are ranked only 2.3th using original mutators and
ranked only 3.6th using all mutators, demonstrating that few manual
efforts will be involved when inspecting the repair results of
\simpr. We looked into the code and found that one potential reason is
the small number of plausible patches even when using all the mutators
since the test suites of the \defectsj subjects are strong enough to
falsify the vast majority of non-genuine patches. To illustrate, the
maximum number of plausible patches that we can have for a bug of
\Closure (the subject with the most candidate patches) is only 26,
which is much smaller than all the other programs (except \Mockito due
to its small number of bugs with plausible patches). We attribute this
to the stronger test suite of \Closure, e.g., \Closure has over 300
contributors and has the largest test suite among all the studied
programs.

\subsection{RQ3: Comparison with the State-of-Art} \label{sec:rq3}
\parabf{Effectiveness} To investigate this question, we compare \simpr with the state-of-the-art \apr
techniques that have been evaluated on \defectsj, including \capGen
\cite{bib:WCWHC18}, \jaid \cite{bib:LPF17}, \acs
\cite{xiong2017precise}, \hdRepair \cite{le2016history}, \xpar
\cite{le2016history} (a reimplementation of PAR
\cite{kim2013automatic}), \nopol \cite{bib:XMDCMDBM17}, \jGenProg
\cite{martinez2015automatic} (a faithful reimplementation of GenProg
\cite{bib:GNFW12} for Java), \jMut \cite{bib:MM16} (a faithful
reimplementation of source-level mutation-based repair \cite{bib:DW10}
for Java), and \jKali \cite{bib:MM16} (a reimplementation of Kali
\cite{bib:QLAR15} for Java). Following
\cite{xiong2017precise, bib:WCWHC18, bib:LPF17}, we obtained the repair
results for prior \apr techniques from their original papers. Note
that the prior \apr studies often target different subsets of \defectsj{} where
the techniques could be applied. As also mentioned in
\cite{bib:LPF17}, despite this lack of information, comparing the fixed
bugs among different tools remains meaningful. This is because the bugs
that are excluded from one study might be considered as being beyond the
capabilities of the corresponding tool/technique. In Table \ref{tab:comp},
the column ``Techs'' lists all the compared techniques. The column ``All
Positions'' presents the number of genuine and non-genuine plausible
patches found when inspecting all the generated plausible patches for
each bug. Similarly, the columns ``Top-10 Positions'' and ``Top-1
Position'' present the number of genuine and non-genuine plausible
patches found when inspecting Top-10 and Top-1 plausible patches. From
the table, we can observe that \simpr can fix the most number of bugs
compared to all the studied techniques. For instance, \simpr can fix
\allgenuines bugs when considering all plausible patches (18 more than
the 2nd best technique, \jaid), fix 41 bugs when considering Top-10
plausible patches (19 more than the 2nd best \capGen),
and fix 23 bugs when considering only the Top-1 plausible
patches (2 more than the 2nd best \capGen). Note that all the
techniques that can rank larger ratio of genuine patches within Top-1
(i.e., \acs and \capGen) used some software learning/mining
information to guide the patch prioritization process, while \simpr
simply used the simplistic Ochiai fault localization formula.

Another interesting observation worth discussion is that \simpr
produces only non-genuine plausible patches for more bugs than the
other techniques. We found several potential reasons. First, \simpr
simply uses Ochiai and does not use any mining or learning information
\cite{xiong2017precise,bib:WCWHC18} for the patch prioritization or
reduction. Second, \simpr is able to explore a large search space
during a short time due to the lightweight bytecode-level patch
generation, while existing techniques usually have to terminate early
due to time constraints. One can argue that this is a low repair
precision \cite{xiong2017precise}. However, our main goal in this work is to
propose a baseline repair technique that does not require any
mining/learning information for both practical application and experimental 
evaluation; also, recently various patch correctness checking
techniques~\cite{xiong2018identifying, tan2016anti} have been
proposed, and can be directly applied to further improve the \simpr
patch validation process. \Comment{In the future, we plan to utilize
  the existing mining and learning techniques
  \cite{xiong2017precise,bib:WCWHC18}.} Furthermore, in this work, we
also manually inspected all the 105 bugs for which \simpr is only able
to produce non-genuine plausible patches. Surprisingly, we observe
that even the non-genuine plausible patches for such bugs can still
provide useful debugging hints. For example, the plausible patches
ranked at the 1st position for 45 bugs share the same methods with the
actual developer patches, i.e., for 43\% cases the non-genuine
plausible patches can directly point out the patch locations for
manual debugging. In the future, we plan to perform user studies with
our industrial collaborators to further investigate the effectiveness
of such non-genuine plausible patches for manual debugging.

\begin{wrapfigure}{r}{0.18\textwidth}
%\begin{figure}
  \includegraphics[scale=0.5]{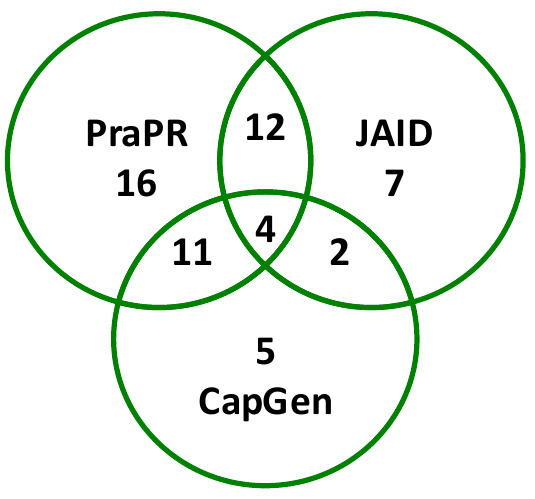}
  \caption{\label{fig:venn} Bugs fixed by most recent \apr tools.}
\end{wrapfigure}

We next manually inspect whether the bugs fixed by \simpr can also be
fixed by the other techniques. Figure \ref{fig:venn} presents the
distribution of the bugs that can be successfully fixed by \simpr and
the two most recent \apr techniques. From the figure, we can observe
that \simpr can fix 16 bugs that have not been fixed by either technique.
Also, the tools are complementary, e.g., \capGen and \jaid each can
fix 5 and 7, resp., bugs that no other tool can fix. \emph{Even when compared
with all the techniques studied in Table \ref{tab:comp}, in total, \simpr
fixes 13 bugs that have not been fixed by any other technique.} Besides \capGen
and \jaid, all the other techniques in Table \ref{tab:comp} are only applicable
for \Chart, \Time, \Lang, and \Math. Interestingly, even on the same 4 programs,
\simpr can still fix 23 bugs that \acs cannot fix, 13 bugs that \hdRepair
cannot fix, 23 bugs that \nopol/\jGenProg/\jKali cannot fix, and 22 bugs
that \jMut cannot fix.

\Comment{We also investigated the efficiency of the compared techniques. Note
that most existing repair techniques cannot exhaustively explore all
the search space, and are only applied within certain time limit
(e.g., 90 minutes limit for \capGen and \hdRepair, while 3
hours for \jGenProg and \nopol), while \simpr has been directly
applied on all the \defectsj bugs exhaustively. Here we further use
some examples to illustrate the efficiency of \simpr. On bug \Lang-6,
the most recent \capGen technique requires 10.9 minutes to explore 216
candidate patches on a machine with 2x Intel Xeon E5-2450 Core CPU
@2.1GHz and 192GB physical memory, while \simpr with all mutators only
needs 1.2 minutes to explore 251 candidate patches on a machine with
a similar configuration, i.e., 10+X speedup. On bug \Closure-31,
another recent advanced \jaid technique takes 991.6 minutes to explore 1,500
candidate patches on a machine with Intel Xeon Processor E5-2630 v2
and 8GB RAM, while \simpr with all mutators only takes 42.8 minutes to
explore all 18,229 patches with a similar configuration, i.e.,
280+X speedup.}

\parabf{Efficiency} We also investigate the efficiency of \simpr
compared to other techniques. \Comment{Note that most existing repair
  techniques cannot exhaustively explore all the search space, and are
  only applied within certain time limit (e.g., 90 minutes limit for
  \capGen and \hdRepair, while 3 hours for \jGenProg and \nopol),
  while \simpr has been directly applied on all the \defectsj bugs
  exhaustively.  Last but not least,} We have executed the most recent
\apr tools \capGen and \jaid on the same hardware platform with \simpr
for having a fair comparison. Table \ref{tab:compTime} shows the detailed time
data on the bugs that \capGen or \jaid can fix. In the table, the columns
1 and 2 present the buggy version information. The columns 3 to 8 present
the number of patches validated and the time cost for \capGen, \jaid,
and \simpr. Finally, the columns 9 and 10 present the speedup achieved by
\simpr over \capGen and \jaid in terms of the time spent on each
validated patch. Note that the highlighted row marks a buggy version
for which the script for running \jaid was unable to download and
initialize.\Comment{ So, it is not known that what the behavior of
  \jaid would look like in fixing this bugs. It is worth mentioning
  that the original paper of \jaid \cite{bib:LPF17} reports that this
  bug is fixed by a patch which is ranked as Top-1. Unfortunately, the
  current implementation of \capGen does not allow us to freely test
  the tool on different benchmark programs that are not listed in the
  original paper \cite{bib:WCWHC18}.}  According to Table
\ref{tab:compTime}, \simpr is 26.1X and 15.7X faster than \capGen and
\jaid{}, respectively. Note that \simpr is still an order of magnitude
faster than \jaid, even though \jaid applies meta-program encoding to compile
a batch of patches at a time. We attribute this substantial speedup to
the fact that \simpr operates completely at the bytecode level; it
does not need any re-compilation and class/test loading from disk for
any patch that it generates. Our experience in using these tools made
clear that how flexible and practical \simpr is compared to the other
tools. Even though the current implementation of \simpr is just a
prototype serving as a proof of concept for the practicality of our
ideas, it is significantly more flexible and easy to use. For example,
both \capGen and \jaid require various configurations to get started,
and are not designed to be used with arbitrary Java projects (they are
tailored for being used with \defectsj programs; setting up them to
work with other Java projects need a considerable amount of manual
work).\Comment{ In the mean time, we could not find a way to fine tune
  \jaid to use JDK 1.7 for the subject programs. This is while most
  \defectsj programs require JDK 1.7 to behave as expected
  \cite{bib:JJE14}.} On the contrary, \simpr besides bringing a
simple, yet effective, idea into the limelight, it offers a clearly
superior engineering contribution. \simpr is a 1-click Maven plugin
that is publicly available on Maven Central Repository, so it is
applicable to \emph{arbitrary} Java project (not just \defectsj and a
subset of IntroClassJava \cite{bib:IntroClassJava}) and even
projects in other JVM languages (Kotlin is already supported). It is
well-documented and follows conventions of Maven plugins in specifying
settings, thereby making it a user-friendly \apr tool.  \Comment{
  compares running time of sinlge-threaded \simpr with the most recent
  \apr tool CapGen. When we were running CapGen we noticed that most of
  the times the tool takes advatage of the \emph{full} computational
  power of our workstation. We inspected its source code and we did
  not find any use of concurrency API of Java. Once we run CapGen
  while setting processor affinity to a specific core, thereby
  enforcing the tool to run like a single-threaded program, the
  running time changes dramatically. Perhaps the underlying library
  (namely, ASTOR \cite{bib:MM16}) uses concurrency API, or JDK 1.8
  (using which CapGen is built) does some optimizations and runs the
  program concurrently whenever it is possible.  Therefore, with the
  current implementation of CapGen a truly fair comparison is not
  possible. Despite that, as the results show, \simpr runs faster than
  CapGen for the most of cases. Furthermore, the bugs that are listed
  in Table \ref{tab:compTime} are the ones that are fixable by CapGen;
  for other programs, CapGen might loop forever or take a very long
  time to finish searching. \simpr, on the other hand, sweeps the
  whole search space for the largest programs of \defectsj, on
  average, within one hour.  }
  
\begin{table}
\begin{adjustbox}{width=1.0\columnwidth}
\begin{tabular}{l|ll|ll|ll}
\hline
&\multicolumn{2}{c|}{All Positions}&\multicolumn{2}{c|}{Top-10 Positions}&\multicolumn{2}{c}{Top-1 Position}\\
Techs&Gen.&Non-gen.&Gen.&Non-gen.&Gen.&Non-gen.\\
\hline
\hline
%SimPR & 37 & 70 & 35 & 72 & 24 & 83\\
\simpr & 43 & 105 & 41 & 107 & 23 & 125 \\
\hline
\capGen & 22 & 3 & 22 & 3 & 21 & 4\\
\jaid & 25 & 6 & 15 & 16 & 9 & 22\\
ACS & 18 & 5 & 18 & 5 & 18 & 5\\
HD-Repair & 16 & N/A & N/A & N/A & 10 & N/A\\
xPAR & 4 & N/A & 4 & N/A & N/A & N/A\\
NOPOL & 5 & 30 & 5 & 30 & 5 & 30\\
jGenProg & 5 & 22 & 5 & 22 & 5 & 22\\
jMutRepair & 4 & 13 & 4 & 13 & 4 & 13\\
jKali & 1 & 21 & 1 & 21 & 1 & 21\\
\hline
\end{tabular}
\end{adjustbox}
\caption{\label{tab:comp} Comparison with state-of-the-art techniques}
\end{table}

\begin{table}
\begin{adjustbox}{width=\columnwidth}
\begin{tabular}{|l|l||ll||ll||ll||rr|}
\hline
&&\multicolumn{2}{c||}{\capGen}&\multicolumn{2}{c||}{\jaid}&\multicolumn{2}{c||}{\simpr}&\multicolumn{2}{c|}{Speedup} \\
\cline{3-10}
Sub.&BugID&\#Patches&Time&\#Patches&Time&\#Patches&Time&\capGen&\jaid \\
\hline
\hline
\Chart&1&458&1496.9&3762&2805&3555&249&47.7X&11X \\
\Chart&8&193&550&N/A&N/A&515&43&30X&N/A \\
\Chart&9&N/A&N/A&5991&4162&1969&81&N/A&17.4X \\
\Chart&11&263&395.84&N/A&N/A&169&35&6.9X&N/A \\
\Chart&24&105&122&2476&904&133&33&4.4X&1.5X \\
\Chart&26&N/A&N/A&2018&2819&11920&705&N/A&24.1X \\
\cellcolor{GrayTwo}\Closure&\cellcolor{GrayTwo}18&\cellcolor{GrayTwo}N/A&\cellcolor{GrayTwo}N/A&\cellcolor{GrayTwo}N/A&\cellcolor{GrayTwo}N/A&\cellcolor{GrayTwo}46270&\cellcolor{GrayTwo}5252&\cellcolor{GrayTwo}N/A&\cellcolor{GrayTwo}N/A \\
\Closure&31&N/A&N/A&14464&96103&30413&3972&N/A&38.5X \\
\Closure&33&N/A&N/A&4484&15109&60314&6414&N/A&31.3X \\
\Closure&40&N/A&N/A&5243&6703&33819&3638&N/A&11.6X \\
\Closure&62&N/A&N/A&7138&7055&436&114&N/A&3.8X \\
\Closure&63&N/A&N/A&7138&7014&436&111&N/A&3.9X \\
\Closure&70&N/A&N/A&2359&3671&35521&3561&N/A&16.7X \\
\Closure&73&N/A&N/A&11472&22647&9365&766&N/A&24.4X \\
\Closure&126&N/A&N/A&4583&35383&28934&3025&N/A&96X \\
\Lang&6&332&996&N/A&N/A&296&132&7.3X&N/A \\
\Lang&26&821&5634&N/A&N/A&1482&52&285X&N/A \\
\Lang&33&N/A&N/A&792&628&31&20&N/A&1.2X \\
\Lang&38&N/A&N/A&1363&546&1968&119&N/A&6.6X \\
\Lang&43&183&5739&N/A&N/A&288&9873&1X&N/A \\
\Lang&45&N/A&N/A&7173&6164&380&35&N/A&9.1X \\
\Lang&51&N/A&N/A&8514&11148&375&31&N/A&15.1X \\
\Lang&55&N/A&N/A&170&204&182&96&N/A&2.4X \\
\Lang&57&2078&2603&N/A&N/A&10&25&0.5X&N/A \\
\Lang&59&623&963&N/A&N/A&137&27&8.5X&N/A \\
\Math&5&590&506&1426&674&381&1491&0.3X&0.1X \\
\Math&30&390&380&N/A&N/A&1109&729&1.5X&N/A \\
\Math&32&N/A&N/A&2997&1910&25574&2543&N/A&6.3X \\
\Math&33&1957&6093&N/A&N/A&5001&922&18X&N/A \\
\Math&50&N/A&N/A&37848&97247&1098&271&N/A&10.3X \\
\Math&53&310&836&2010&971&250&160&4X&0.8X \\
\Math&57&194&1989&N/A&N/A&502&287&17X&N/A \\
\Math&58&508&1356&N/A&N/A&9002&2994&7.5X&N/A \\
\Math&59&51&518&N/A&N/A&3370&445&76X&N/A \\
\Math&63&168&443&N/A&N/A&135&45&7.5X&N/A \\
\Math&65&1828&6058&N/A&N/A&6098&312&65X&N/A \\
\Math&70&135&151&N/A&N/A&223&34&7.3X&N/A \\
\Math&75&203&202&N/A&N/A&557&40&13.9X&N/A \\
\Math&80&1279&6966&9526&8900&18725&1318&71X&12.9X \\
\Math&82&N/A&N/A&1707&1874&2721&108&N/A&28X \\
\Math&85&247&5540&2922&3905&1599&489&82.5X&4.7X \\
\hline
\textbf{Avg.}&&587.1&2251.7&6149&14106.1&8421&1234.1&26.1X&15.7X\\
\hline
\end{tabular}
\end{adjustbox}
\caption{\label{tab:compTime} Comparison of timing of \simpr with
reproduced timing of \capGen and \jaid{} on the bugs that they can fix}
\end{table}

\subsection{Threats to Validity}

\parabf{Threats to Internal Validity} are mainly concerned with the
uncontrolled factors that may also affect the experimental results.
The main threat to internal validity for this work lies in the
potential faults during the implementation of \simpr. To reduce the
threat, we implemented \simpr under state-of-the-art libraries and
frameworks, such as the ASM bytecode manipulation framework and the
\pit mutation testing engine. We also carefully reviewed all our
scripts and code to detect potential issues. To further reduce this
threat, we plan to release our implementation as an open-source project
and encourage developers/researchers to contribute.

\parabf{Threats to External Validity} are mainly concerned with
whether the experimental findings from the used subject systems can
generalize to other projects. To reduce these threats, to our knowledge, we performed the first repair study on all the \djbug real-world bugs from
the widely used \defectsj benchmark. To further reduce the threats, we
plan to study \simpr on other debugging
benchmarks~\cite{saha2017elixir, long2017automatic}.

\parabf{Threats to Construct Validity} are mainly concerned with
whether the metrics used in our experimental study are well-designed
and practical. To reduce these threats, we used the widely used
metrics in automated program repair research, such as the number of
genuine patches, the number of plausible patches, actual time cost,
and so on. We further evaluated the usefulness of plausible patches in
helping with manual debugging. To further reduce these threats, we
plan to apply \simpr to the daily development of our industry
collaborators and get feedbacks from real-world developers.

\section{Conclusion}\label{sec:ConclusionFW}
In this work, we propose a practical approach to automatic
program repair, named \simpr, based on bytecode-level mutation-like
patch generation. We have implemented \simpr as a practical program
repair tool based on the state-of-the-art mutation engine \pit and ASM
bytecode manipulation framework. The experimental results on the
widely used \defectsj benchmark show that \simpr, using all of its mutators,
can generate genuine patches for \allgenuines \defectsj bugs,
significantly outperforming state-of-the-art \apr techniques, while being
an order of magnitude faster. \simpr is a simplistic technique that
does not require any learning/mining and can easily be used as a
baseline for future repair techniques.

\bibliographystyle{abbrv}
\bibliography{bibdb}

\end{document}